\documentclass[letterpaper]{article}
\usepackage[margin=1in]{geometry}
\usepackage[square,numbers,sort&compress]{natbib}
\usepackage[english]{babel}
\usepackage{color}
\usepackage[small]{titlesec}
\usepackage{rotating}
\usepackage{amsmath}
\usepackage{amssymb}
\usepackage{amsfonts}
\usepackage{bm}
\usepackage{mathrsfs}
\usepackage{supertabular}
\usepackage{graphicx}
\usepackage{subcaption}
\usepackage{ulem}

\usepackage{fancyhdr}
\usepackage{totpages}
\title{Dynamics of spheroids in pressure driven flows of shear thinning fluids}

\author{Vishal Anand and Vivek Narsimhan \footnote{To whom the correspondence should be addressed. \href{mailto:vnarsim@purdue.edu}{\texttt{vnarsim@purdue.edu}}}; \\
\textit{Davidson School of Chemical Engineering, Purdue University,}\\ \textit{West Lafayette, Indiana 47907, USA}}

\usepackage{hyperref}
\hypersetup{
    colorlinks=true,
    citecolor=magenta,
    urlcolor=cyan,
    breaklinks=true
}

\begin{document}

\maketitle

\begin{abstract}
 Particles in inertialess flows of shear thinning fluids are a model representation for several systems in biology, ecology, and microfluidics. In this paper, we analyse the motion of a spheroid in a pressure driven flow of a shear thinning fluid. The shear thinning rheology is characterised by the Carreau model. We use a combination of perturbative techniques and the reciprocal theorem to delineate the kinematics of prolate and oblate spheroids. There are two perturbative strategies adopted, one near the zero shear Newtonian plateau and the other near the infinite shear Newtonian plateau. In both limits, we find that a reduction in effective viscosity decreases the spheroid’s rotational time period in pressure driven flows. The extent to which shear thinning alters the kinematics is a function of the particle shape. For a prolate particle, the effect of shear thinning is most prominent when the spheroid projector is aligned in the direction of the velocity gradient, while for an oblate particle the effect is most prominent when the projector is aligned along the flow direction.  Lastly, we compare the tumbling behavior of spheroids in pressure driven flow to those in simple shear flow. While the time period decreases monotonically with Carreau number for pressure driven flows, the trend is non monotonic for shear flows where time period first increases at low Carreau number and then decreases at high Carreau numbers. Shear thinning does not resolve the degeneracy of Jefferey's orbits.
\end{abstract}

\section{Introduction}
\label{sec:intro}

Rigid, orientable particles experience a bevy of interesting phenomena when placed in highly viscous, structureless fluids \cite{Purcell_AJP_2015}.  For example, a single rod sediments at the same initial orientation in such fluids, while a single rod tumbles in a periodic orbit (Jeffrey orbit) in shear flow.  The reason for these behaviors arise from the symmetry, linearity, and reversibility of the Stokes equations \cite{HB83,Leal2007,KimKarilla2005}.  When the assumptions of the Stokes equations break down – i.e., when fluid inertia is present or the fluid exhibits non-Newtonian rheology –the above behavior may no longer be valid \textcolor{red}{ \cite{lealadvanced} }.

By now, there are plenty of studies that investigate how different particle shapes (e.g., fibers, sheets, spheres, spheroids) move under sedimentation, linear flows, and quadratic flows when the fluid has small but non-zero inertia.  For example, a fibre-like particle was analysed in shear flow of a Newtonian fluid with weak inertia using a slender body approximation and the reciprocal theorem \cite{SK06}. It was shown that the fibre drifts towards the velocity-shear gradient plane due to inertia and then stops rotating altogether above a critical Reynolds number. A similar analysis was performed for spheroids in simple shear flow \cite{DMS16}. The analysis was later extended to general linear flows, which are a combination of extensional and rotational flows \cite{MS18}. The analyses showed that both fluid and particle inertia forced the prolate particle to tumble in velocity -velocity gradient plane, while the oblate particle was forced into a log rolling motion. A linear stability analysis validated the stability of these orientations for prolate and oblate particles \cite{ECLAM15,ECLAM15_A}.

When the suspending fluid is no longer Newtonian, the fluid can experience phenomena such as normal stress differences, shear thinning, extensional thickening, and/or time-dependent viscoelasticity.  Each of these effects play a major role in altering the motion of rigid, orientable particles.  We will not provide a complete survey on the microhydrodynamics of such particles in non-Newtonian fluids – instead, we refer the readers to treatises \cite{DAvino_Review_JNNFM,AGM17} and provide a brief summary here.  Generally, much of the effort on the past few decades has focused on the effect of normal stresses on particle motion \cite{Shaqfeh_Eisntein_Viscosity,Shaqfeh_Viscoelastic_PRF,Shaqfeh_Viscoelastic_Shear}.  Normal stresses give rise to two different phenomena in low Reynolds number flows -- namely cross stream migration and steady state orientation. Cross-stream migration refers to the lift forces on particles flowing in channel.  This phenomenon occurs when there is a gradient in normal stresses, and thus is generally observed when there is a gradient in strain rates – i.e., a quadratic flow \cite{Wang_Narsimhan_POF}.  During sedimentation and/or shear flow, normal stresses can also give rise to a hydrodynamic torque on orientable particles.  This situation leads to a stable orientation \cite{Wang_Narsimhan_POF,Brunn_Review_Paper}. The analytical solution for the sedimentation of spheroids in a quiescent background of a second order fluid was furnished by \cite{Kim86}, while those for linear flows by \cite{B77}. Recently, these studies were extended by \cite{Tai_Wang_Narsimhan_JFM}, who furnished analytical formulae for the polymeric force and torque on a spheroid, in quadratic flows of second order fluids in the limit of small Weissenberg number. A recent survey of the research into viscoelastic microhydrodynamics is furnished in \cite{AGM17}.

 There has been interest in the microhydrodynamics of particles in fluids with variable viscosity, where the variation in viscosity is either spatial \cite{GVG_2022,Vaseem_Elfring_Viscosity,Datt_Elfring_Viscosity_Gradient} or due to the shear thinning rheology of the fluid \cite{Datta_Elfring_JNNFM_ShearThinning,Elfring_Main}. Out of the two, the focus in this paper is the motion of spheroids in shear thinning fluids; such systems are prevalent in biology, microfluidics, and ecology \cite{Elfring_Local_Global,Elfring_Reciprocal_JFM}.  There particularly has been work done for microswimming (active particles)\cite{VDPP22}, due to the intuitive but naive speculation that a reduction in viscosity afforded by shear thinning allows the swimmer to swim faster without expending as much as energy \cite{Lauga_Elfring_2015,Elfring_Reciprocal_JFM}. For spherical squirmers, an asymptotic analysis revealed both enhancement and reduction of swimming velocity depending on the surface actuation of swimmers \cite{Datt_Charu_Squirming_JFM_2015,Swimming_Efficiency_PRE_2017}. There are shear rates at which the swimming velocity gets optimized \cite{Swimming_Efficiency_PRE_2017}. Thus, the naive hypothesis that shear thinning enhances the swimming speed is faulty and must be used with caution \cite{Swimming_Efficiency_PRE_2017,Datt_Charu_Squirming_JFM_2015}. Very recently, the swimming characteristics of Purcell's swimmer in a shear thinning fluid was analysed where it was revealed that unequal arm rotation rates induce a net vertical displacement, which is not present in Newtonian fluids \cite{Pak_PurcellSwimmer}.

Despite the recent progress on particle dynamics in shear thinning fluids, it is clear that this area of research is relatively underexplored with many challenging problems remaining to be investigated.  For instance, most of the asymptotic analyses of particle dynamics explore the Carreau model of fluid rheology, performing a perturbation expansion in the small Carreau number (i.e., low shear rate) limit. We still do not know what happens when the Carreau number is not small. Moreover, for passive particles, the analysis has been restricted to spheres, with the notable exception of \cite{Elfring_Main}. Other particle shapes, like oblate spheroids, have not been investigated with respect to their interaction with shear thinning flows. Finally, we understand that microhydrodynamics of particles during sedimentation \cite{Datta_Elfring_JNNFM_ShearThinning} and linear shear flows \cite{Elfring_Main} of shear thinning fluids has been analysed, but more complicated flows like pressure driven flows have not been discussed. In this context, we propose to answer the following research questions in this paper:  \textit{How does the orientational kinematics of prolate and oblate spheroids vary in pressure driven flows of shear thinning fluids in both the small and large Carreau number limits?   How is this behavior different than in the simple shear flow case?} 

The paper is organized as follows. First we introduce the problem formally in Sec.~\ref{sec:Problem_Statement}. The definition of prolate and oblate spheroids are discussed here as well as the fluid rheological model (Carreau model). Next, in Sec.~\ref{sec:theory}, we show how the pressure driven flow is altered due to shear thinning alone, without the presence of any particle. This section also introduces the particle into this flow field, and uses the reciprocal theorem to quantify the shear thinning correction to the particle kinematics. In the next section, we discuss the algorithm of the code (Sec.~\ref{sec:code}), which we use to obtain the results in Sec.~\ref{sec:results_discussion}. In  Sec.~\ref{sec:LinearFlow}, we present some results pertaining to the tumbling time period of spheroids in simple shear flows in both the small and large Carreau number regimes and discuss the difference between the trends observed in shear flows vis-a-vis those observed in pressure driven flows. A discussion and conclusion follow in Sec.~\ref{sec:conclusion}.

\section{Problem Statement} 
\label{sec:Problem_Statement}

\subsection{Problem geometry and particle definition}
\label{sec:part_def}
\begin{figure}[t]
\centering
\subfloat[]{\includegraphics[width=0.45\linewidth]{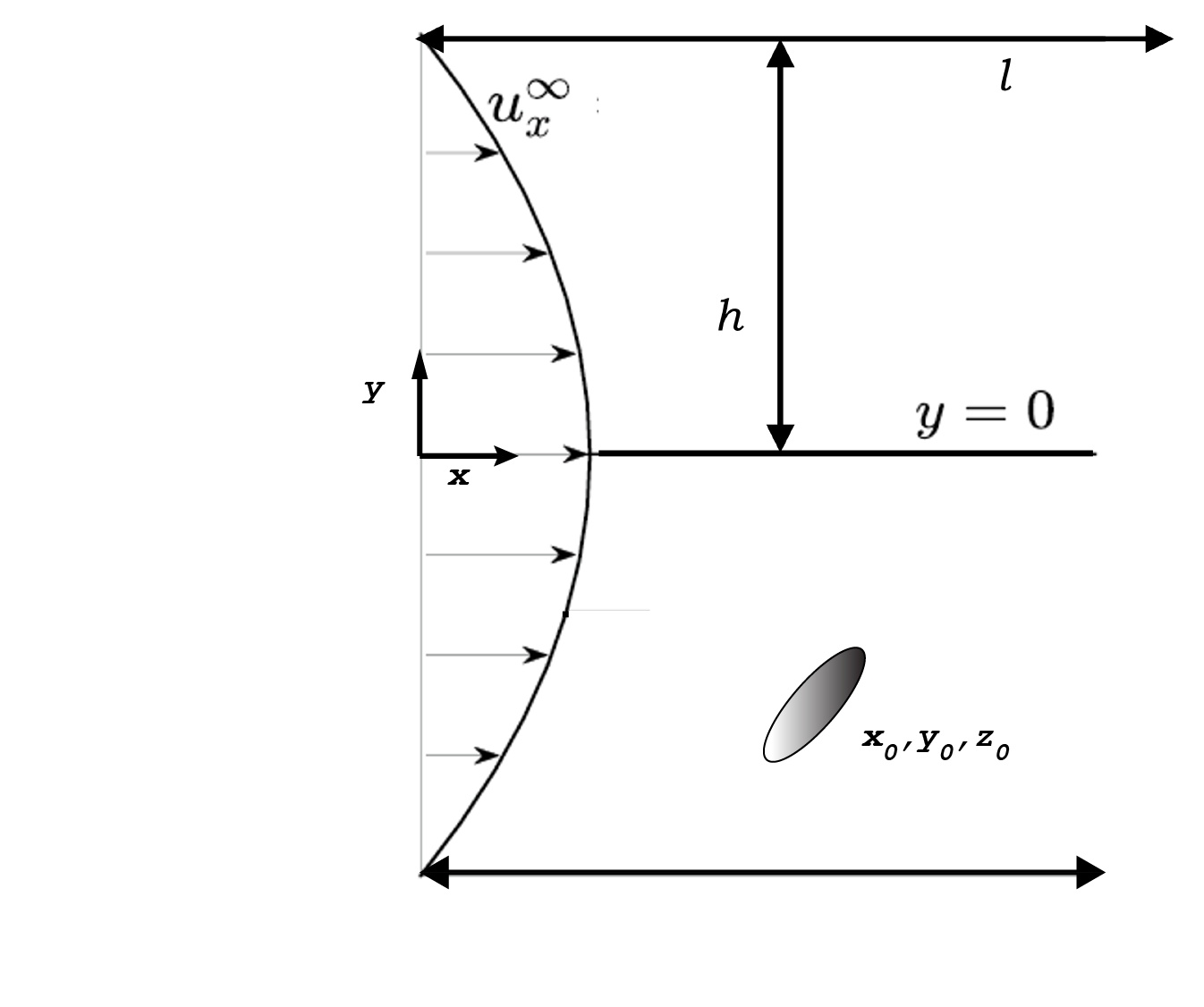}}
\hfill
\subfloat[]{\includegraphics[width=0.45\linewidth]{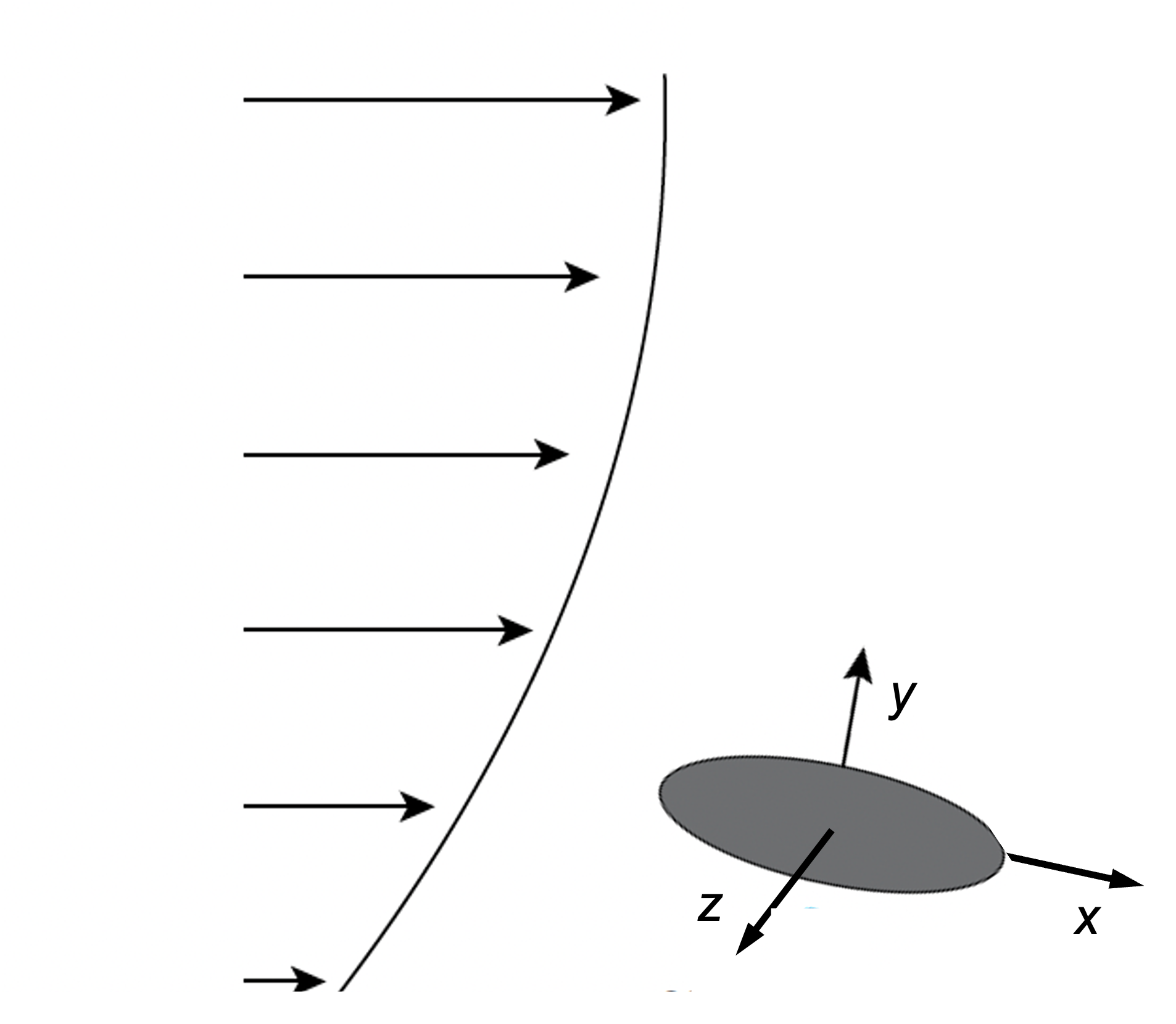}}
\caption{System geometry in a shear thinning fluid. (a) Lab coordinate system.  The channel has half width $h$ and length $l$, and the coordinate axes are placed at the center of the channel inlet. A spheroid is initially placed at location $(x_0,y_0,z_0)$ and its position and orientation is tracked over time. The dimensions are not to scale. (b) Particle coordinate system.  When solving for the particle’s rigid body motion, we assume the particle is in an unbound fluid with a velocity $u_i^{\infty}$ given by the flow in the channel.  The coordinates align with the particle’s semi-axes and the origin is the particle’s center of mass.}
\label{fig:Schematic_1}
\end{figure}

The system under investigation is a spheroid in a pressure driven flow of a shear thinning fluid. The flow is steady and inertialess, and the particle is neutrally buoyant and passive. Figure \ref{fig:Schematic_1}a illustrates the geometry of the system.  We investigate a slit-like channel of length $l$, height $2h$, and infinite width.  The coordinate system is positioned such that the $x$-axis aligns with the midplane of the channel while the $y$-axis aligns with the height of the channel.  A pressure $p=\Delta P$ is imposed at the inlet of the channel $x=0$, while the outlet pressure at $x =l$ is $p=0$.

The spheroid will start at a position $(x_0, y_0, z_0)$ in the channel.  The lengths of the three semi-axes are $a, b,$ and $c$, where $b = c$.  The unequal axis -- i.e., the direction of the $a$-axis –- is known as the projector.  A prolate spheroid has its unequal axis the longest ($a > b$), while an oblate spheroid has its unequal axis the shortest ($a < b$).  Apart from the semi-axes ($a,b,c$), it is also possible to parameterize the spheroid shape using two other quantities $R$ and $A_R$. $R$ is the radius of the equivalent sphere with the same volume as the spheroid, which means $V=\frac{4}{3} \pi R^{3} =\frac{4}{3}\pi a b c$. The quantity $A_R$ is the ratio of the projector axis to the other two axes: $A_R =a/b$. By this definition,  prolate spheroids have $A_R > 1$, while  oblate spheroids have $A_R < 1$. The two systems of particle parameterization are connected by

\begin{equation}
\label{eq:part_def_2}
a=R A_{R}^{2 / 3}, \quad b=c= R A_R^{-1/3} 
\end{equation}

\begin{figure}[t]
\centering
\subfloat[]{\includegraphics[width=0.45\linewidth]{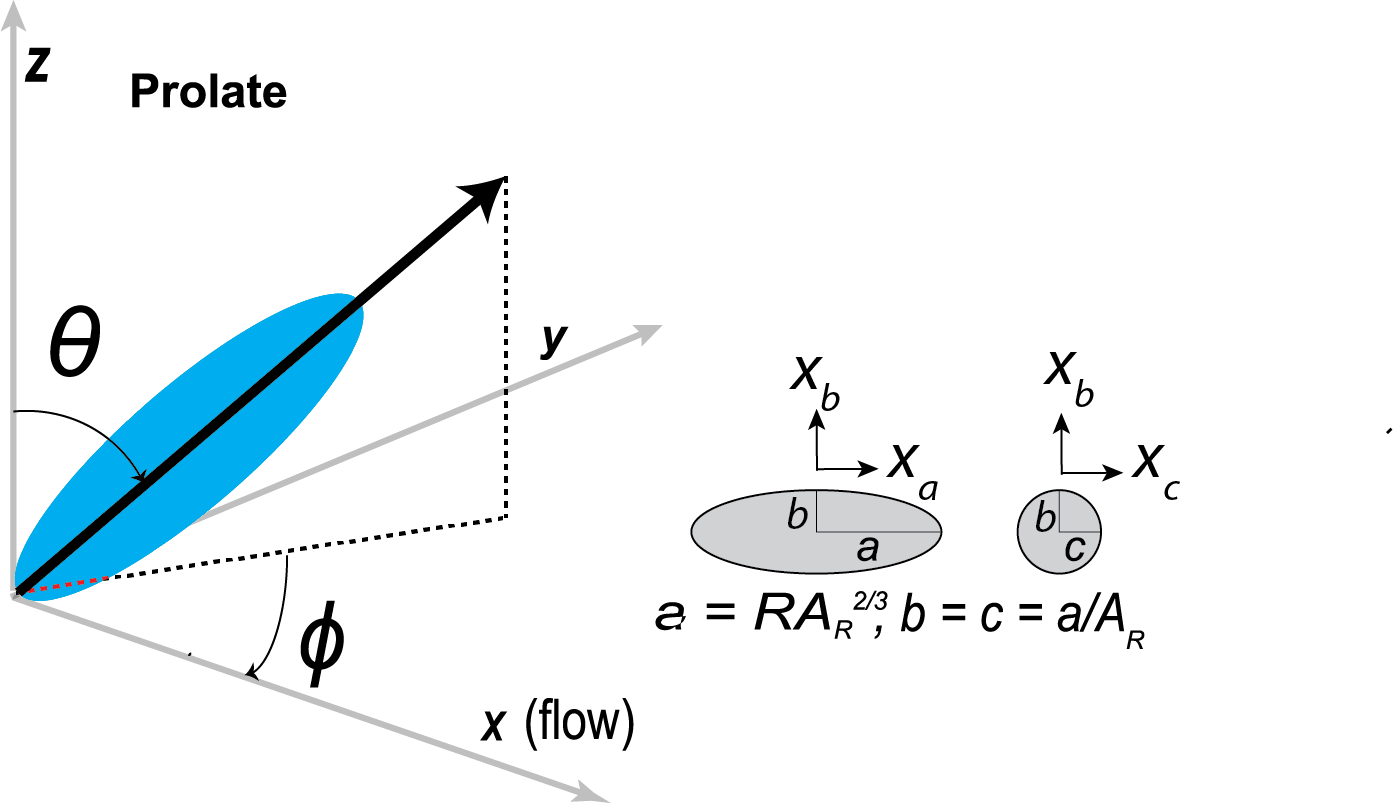}}
\hfill
\subfloat[]{\includegraphics[width=0.45\linewidth]{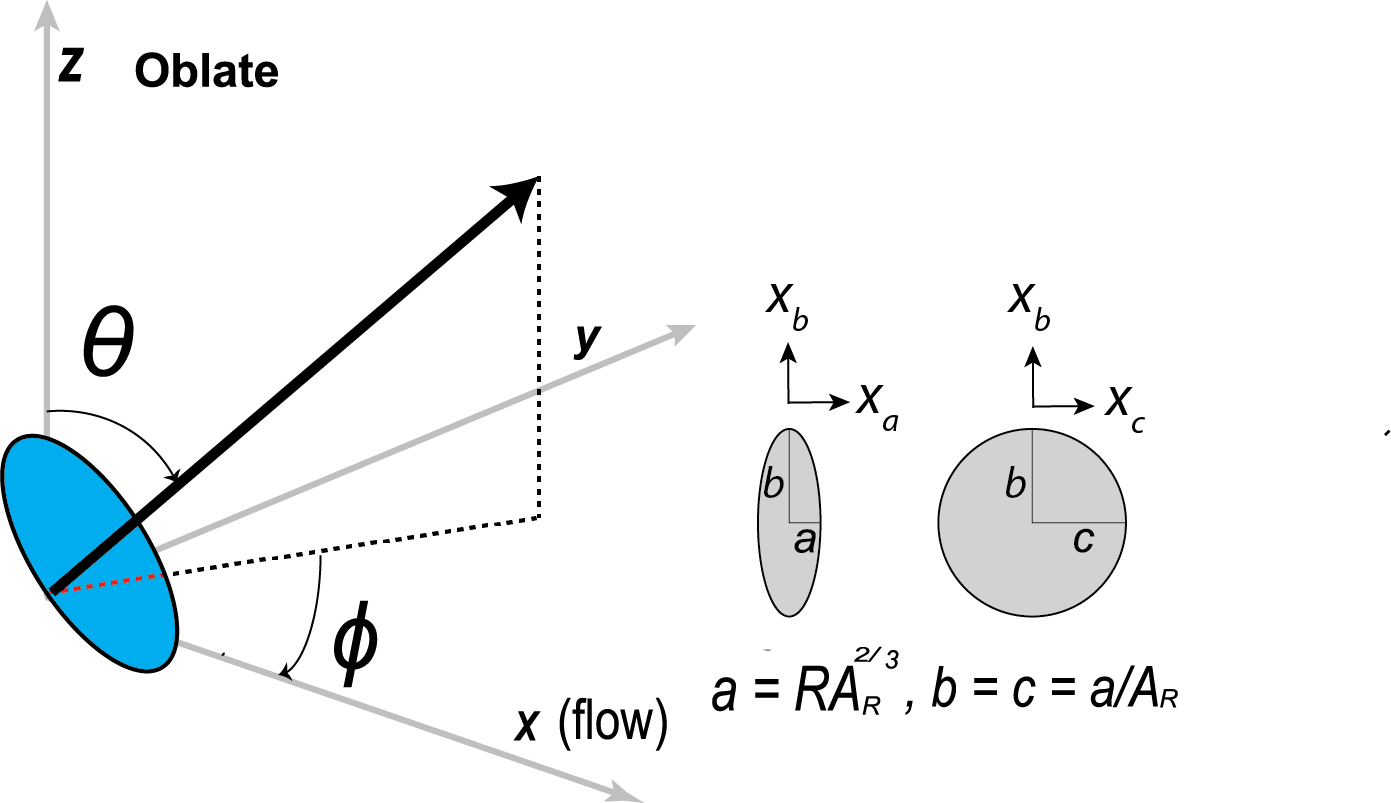}}
\caption{Definition of the particle shape and orientation for both (a) prolate and (b) oblate spheroids. The projector makes an angle $\phi$ (azimuth) in the $x-y$ plane with flow direction $x$, and an angle $\theta$ (polar) with the $z$ direction. For both prolate and oblate spheroids, the relationship is also shown between the two different systems of parameterizations, based on $a,b,c$ and on $R$ and $A_R$ respectively.}
\label{fig:Particle_def}
\end{figure}

The orientation of the spheroid is characterized by the ordered pair ($\theta, \phi$). Here $\theta \in (0,\pi)$ is the angle from the $z$-axis, also denoted as the co-latitude angle or the polar angle, while $\phi \in (0,2\pi)$ is the angle in $x-y$ plane from the $x$-axis, also known as the azimuth angle.  Figure \ref{fig:Particle_def} summarizes the geometrical definitions for the spheroids.

We will solve for the motion of the spheroid in the limit when the particle size is much smaller than the channel height – i.e, $R \ll h$ .  In this situation, one can neglect the hydrodynamic interaction with the wall and treat the particle as if it were in an unbound fluid with a background velocity $u_i^{\infty}$ given by the flow field in the channel.  This approximation incurs an error of $O( (R/h)^3)$ for the rigid body motion (see \citet[Ch.~12]{KimKarilla2005}).  Thus, in this problem, we will often switch between two coordinate systems.  When solving the translational and rotational velocity of the particle, we will use Figure \ref{fig:Schematic_1}b and treat the particle in an unbound medium with the origin at the particle’s center of mass and local coordinate system aligned with the particle's semi-axes.  Once we determine these quantities, we will go back to the channel coordinates (Figure \ref{fig:Schematic_1}a) and update the particle position and orientation over a time step $\Delta t$.  The process will repeat until we track the particle motion for a given time interval.  In this context, the microhydrodynamic problem encountered here may be considered as a case of fluid structure interaction problem \cite{AC18b,AC19a,AMC20,ADC18,VAN22,AC20}; the flow field exerts pressure/force on the structure leading to the motion (or deformation, if the solid is soft) of the particle. The next subsection describes the rheological model for the fluid in the channel.




\subsection{Fluid rheology}
\subsubsection{Carreau-Yasuda model}
\label{sec:fluid_rheo}
The background fluid is complex \cite{BAH87,Anand2016EffectFormulation,Anand2015EntropyFluid}. Specifically, the fluid is shear-thinning – i.e., the fluid exhibits a reduction in the apparent viscosity with increase in the applied shear stress.  Such fluids are also called pseudoplastics and common examples include polymers, blood, and ketchup among others \cite{CR08}(see also \cite{ACJHT_2019,Anand2014}).  The constitutive equation for the stress is the same as a Newtonian fluid, except that the viscosity is a function of the strain rate:

\begin{equation} \label{eq:Carreau}
\tau_{ij} = \eta(\dot{\gamma}) \dot{\gamma}_{ij} – p \delta_{ij}; 
\end{equation}

In the above equation, $p$ is the pressure, $\dot{\gamma}_{ij} = \frac{\partial u_i}{\partial x_j} + \frac{\partial u_j}{\partial x_i}$ is twice the rate of strain tensor, and $\eta$ is the viscosity.  The viscosity is a function of the magnitude of strain rate, given by $\dot{\gamma} = \sqrt{\frac{1}{2} \dot{\gamma}_{ij} \dot{\gamma}_{ij} }$, where Einstein convention is assumed.

Rheological models for shear thinning viscosity include the power law model of Ostwald–de Waele, the Herschel-Bulkley model, and the Carreau-Yasuda model \cite{BAH87}.  Here, we employ the Carreau-Yasuda model, mainly because this model captures both Newtonian plateaus at the beginning and the end of shear thinning regimes.  The constitutive equation for the Carreau-Yasuda model is given as:
\begin{equation}
\label{eq:consti}
    \eta =\eta_{\infty}+(\eta_{0}-\eta_{\infty})[1+\lambda_t^2\dot{\gamma}^2]^{\frac{n-1}{2}}
\end{equation}

In this model, the fluid behaves Newtonian at low and high shear rates, with a power-law region in-between.  The quantities $\eta_{0}$ and $\eta_{\infty}$ are the zero shear rate and infinite shear rate viscosities, respectively.  The index $n < 1$ determines the rate of decay in the power-law region, while $\lambda_t$ is the time constant that determines the strain rate at which the power-law region occurs. 

\subsubsection{Non-dimensional form of Carreau-Yasuda model}

The dimensionless form of the Carreau-Yasuda model is given by:

\begin{equation}
\label{eq:consti_dimless}
    \eta =1-\epsilon+\epsilon[1+Cu^2\dot{\gamma}^2]^{\frac{n-1}{2}}
\end{equation}
where the viscosity is rendered dimensionless by $\eta_{0}$ and the shear rate $\dot{\gamma}$ is rendered dimensionless by a characteristic shear rate $\dot{\gamma}_c$.  The characteristic shear rate is determined by the problem under consideration (such as shear flow or pressure driven flow).  There are two new dimensionless numbers introduced above.  The relative viscosity drop
\begin{equation}
    \epsilon =1-\beta =\frac{\eta_0 - \eta_{\infty}}{\eta_{0}}
\end{equation}
represents the fractional drop in viscosity between the zero and infinite shear rate limits.  The Carreau number 
\begin{equation}
    Cu = \lambda_t \dot{\gamma_c}
\end{equation}
describes the shear-thinning regime at the characteristic strain rate $\dot{\gamma}_c$.  When $Cu \ll 1$ or $Cu \gg 1$, the characteristic shear rate is located near the Newtonian plateaus, while for $Cu \sim O(1)$ the characteristic shear rate is in the power-law regime.  A summary of the dimensionless numbers in the Carreau model is shown in Table \ref{tab:rheo}. For illustration, typical values of the rheological parameters of Carreau model pertaining to aqueous solution of $0.3\%$ xanthan gum (molar mass = $933.748$ g/mol) are $n =  0.402$ , $\lambda_t =239$ s and $\beta =1.35 \times 10^{-4}$ \cite{Boyko_Stone_JFM} (see also \cite{Pipe_McKinley_2008}).

\begin{table}[]
\centering
\caption{Summary of dimensionless rheological parameters for Carreau shear thinning fluid.}
\label{tab:rheo}
\begin{tabular}{|l|l|l|l|}
\hline
Dimensionless rheological       & Definition & Relevance & Values  \\
parameter & & & \\ \hline
$Cu$      & $Cu = \lambda_t \dot{\gamma_c}$ & ${\text{ Ratio of characteristic strain rate  }}$ & $0\leq Cu< \infty$ \\ & & {\text{  and  critical strain rate}}& \\ \hline
$n$  & Power law index    & \text{Slope of shear thinning region}   & $0<n<1$ \\   & & &  \\ \hline
$\beta$    & $\beta =\frac{\eta_{\infty}}{\eta_0}$ &   \text{Ratio between infinite shear viscosity}   & $\beta \ll 1$   \\  & & \text{and zero shear viscosity} &     \\ \hline
$\epsilon $  & $1-\beta$              &  \text{Relative viscosity drop}      & $\epsilon \sim 1$        \\     &  & & \\ \hline
\end{tabular}
\end{table}

\subsubsection{Perturbation expansion for constitutive equation}
The rheological model under consideration (Eq.~\eqref{eq:consti_dimless}) is nonlinear in $\dot{\gamma}, Cu$ and $n$. In anticipation of the subsequent microhydrodynamical study, we will express the viscosity in the following form \cite{lealadvanced}:
\begin{equation}
\label{eq:std_perturb}
    \eta =\mu+\delta A{[\dot{\gamma}]},
\end{equation}
where $\mu$ is a (constant) Newtonian viscosity, $A$ is a function of the shear rate, and $\delta$ is a small parameter such that when $\delta \to 0$, the Newtonian limit is recovered. We will examine two different perturbative limits.

\paragraph{Small Carreau number:}  When $Cu^2\dot{\gamma}^2 \ll 1$, one can Taylor expand the viscosity around the zero shear rate plateau.  The small parameter is $\delta = Cu^2$ in the perturbation expansion. We find $\mu = 1$ and $A[\dot{\gamma}] = \frac{1}{2}\epsilon (n-1) \dot{\gamma}^2$, which yields:
\begin{align}
\label{eq:consti_low_Cu}
\eta =1+\frac{1}{2}\epsilon (n-1)Cu^2\dot{\gamma}^2 + O(Cu^4),
\end{align}

\paragraph{Large Carreau number:}  When $Cu^2\dot{\gamma}^2 \gg 1$, one can Taylor expand the viscosity around its infinite shear rate plateau.  Here, the small parameter is $\delta = Cu^{(n-1)}$ in the perturbation expansion.  We find $\mu = \beta$ and $A[\dot{\gamma}] = \epsilon \dot{\gamma}^{n-1}$, which yields:
\begin{equation}
\label{eq:consti_High_Cu}
\eta =\beta+Cu^{(n-1)}\epsilon \dot{\gamma}^{n-1} + O(Cu^{2(n-1)})
\end{equation}

\begin{figure}
  \centering
  \includegraphics[width=0.75\linewidth]{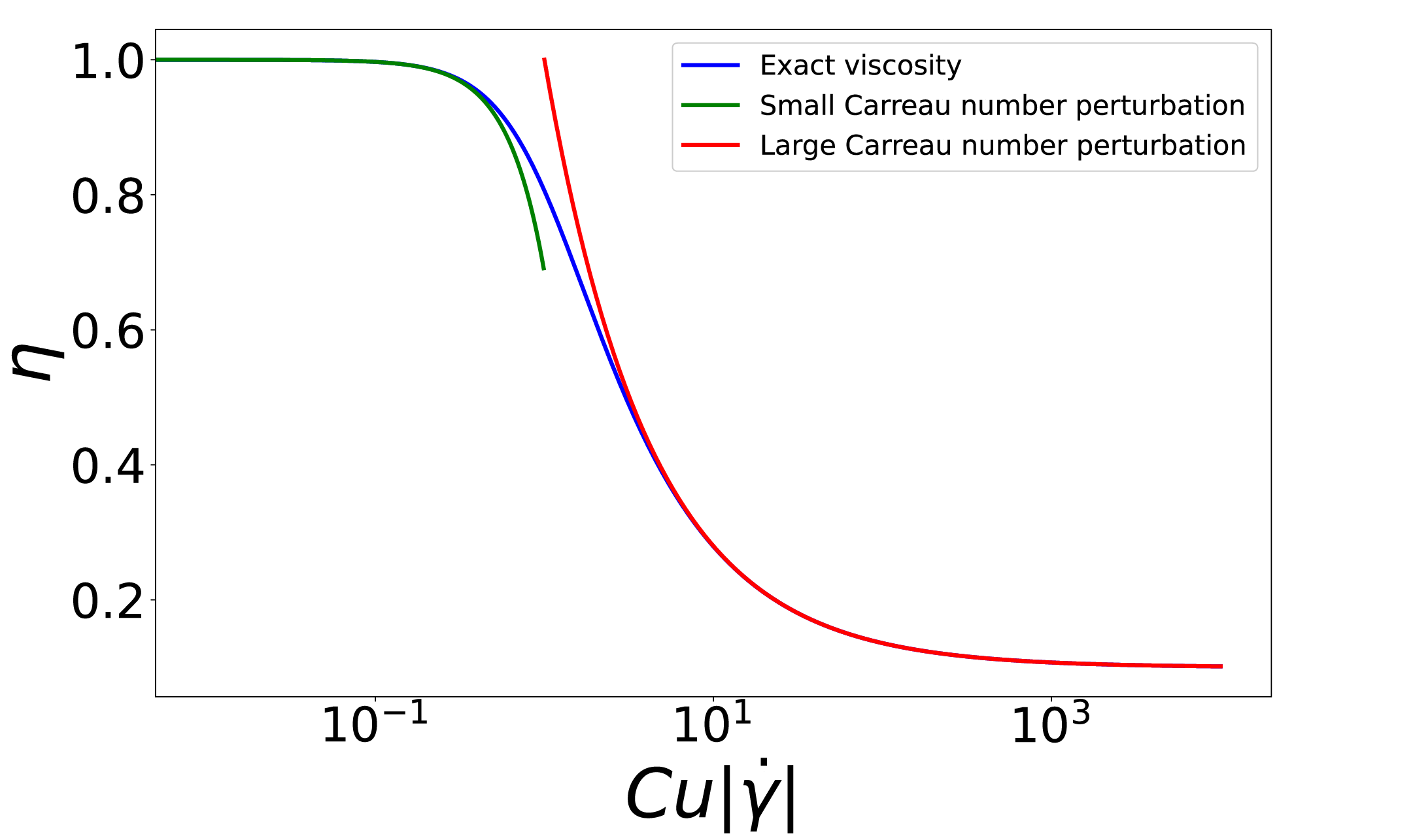}
  \caption{Viscosity of Carreau fluid for $n=0.3$ and $\epsilon =0.9$. The full rheological equation (Eq.~\eqref{eq:consti_dimless}) is compared against the perturbation results in the small Carreau number limit (Eq.~\eqref{eq:consti_low_Cu}) and the large Carreau number limit (Eq.~\eqref{eq:consti_High_Cu}).   }
\label{Fig:Viscosity_Composite}
\end{figure}

In Fig.~\ref{Fig:Viscosity_Composite}, we compare the full rheological equation (Eq.~\eqref{eq:consti_dimless}) to the asymptotic limits discussed above.  Overall, we see that the small Carreau number limit does a reasonable job capturing the rheology for $Cu^2 < 0.1$, while the large Carreau number limit does a reasonable job for $Cu > 10$.

\section{Theory} \label{sec:theory}
\subsection{Overview of steps and non-dimensionalization}
The aim of this work is to analyse the microhydrodynamics of spheroids in shear thinning fluids. We will segregate our analysis into two parts.  First, we will calculate the background flow field of the shear thinning fluid without the particle.  This analysis will be followed by an investigation of a spheroid in this flow field.  The reciprocal theorem will be used to obtain corrections to the rigid body motion, in the limits of small and large Carreau numbers.

From here on out, we will write results in non-dimensional form.  All distances will be scaled by the channel half height $h$.  The viscosity will be scaled by the zero-shear rate viscosity $\eta_0$, while the stresses will be scaled by $\tau_c = \frac{h \Delta P}{2 l}$.  The shear rate will be scaled by $\dot{\gamma}_c = \tau_c/\eta_0 = \frac{h \Delta P}{2 \eta_0 l}$, the time will be scaled by $\dot{\gamma}_c^{-1}$, while the velocities will be scaled by $V_c = \dot{\gamma}_c h$.  The Carreau number for this geometry will be defined as:

\begin{equation}
    Cu = \frac{\lambda_t h \Delta P}{2\eta_0 l}
\end{equation}

\subsection{Background flow field} \label{sec:background_flow}
In this subsection, we will use the coordinate system in Figure \ref{fig:Schematic_1} (a) and solve for the flow field in the channel.  The quantities in this subsection will have a superscript ``$\infty$'' indicating that they are solved in the absence of a particle.  

When the flow is steady, inertialess, and incompressible, the momentum and continuity equations state that $\frac{\partial \tau_{ij}^{\infty}}{\partial x_j} = 0$ and $\frac{\partial u_i^{\infty}}{\partial x_i} = 0$ (Einstein convention assumed).  The flow is unidirectional (i.e., $u_x^{\infty} = u_x^{\infty}(y)$ only), which simplify the equations considerably.  The differential equation for the flow field is:
\begin{equation}
\frac{\partial \tau_{xy}^{\infty}}{\partial y}=-2
\end{equation}
where $\tau_{xy}^{\infty}$ is given by the Carreau model (Eqns.~\eqref{eq:Carreau} and \eqref{eq:consti_dimless}) with local shear rate $\dot{\gamma}^{\infty} =  \frac{\partial{u_x^{\infty}}}{\partial y}$.  The differential equation is subject to the boundary conditions $u^{\infty}_x = 0$ at the top wall ($y =h$) and $\frac{\partial u_x^{\infty}}{\partial y} = 0$ at the center plane ($y = 0$). 

\begin{figure}[t]
\centering
\subfloat[]{\includegraphics[width=0.45\linewidth]{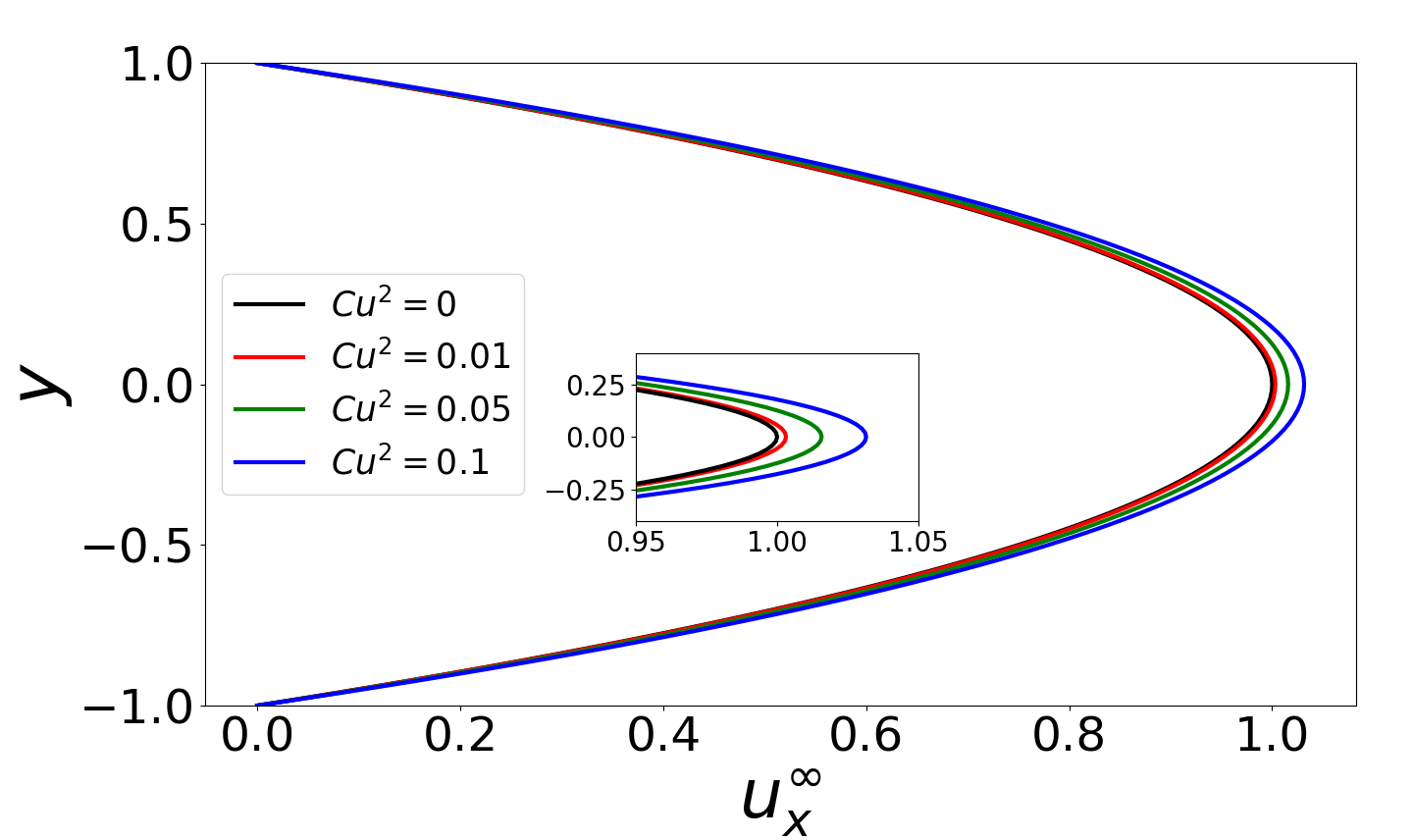}}
\hfill
\subfloat[]{\includegraphics[width=0.45\linewidth]{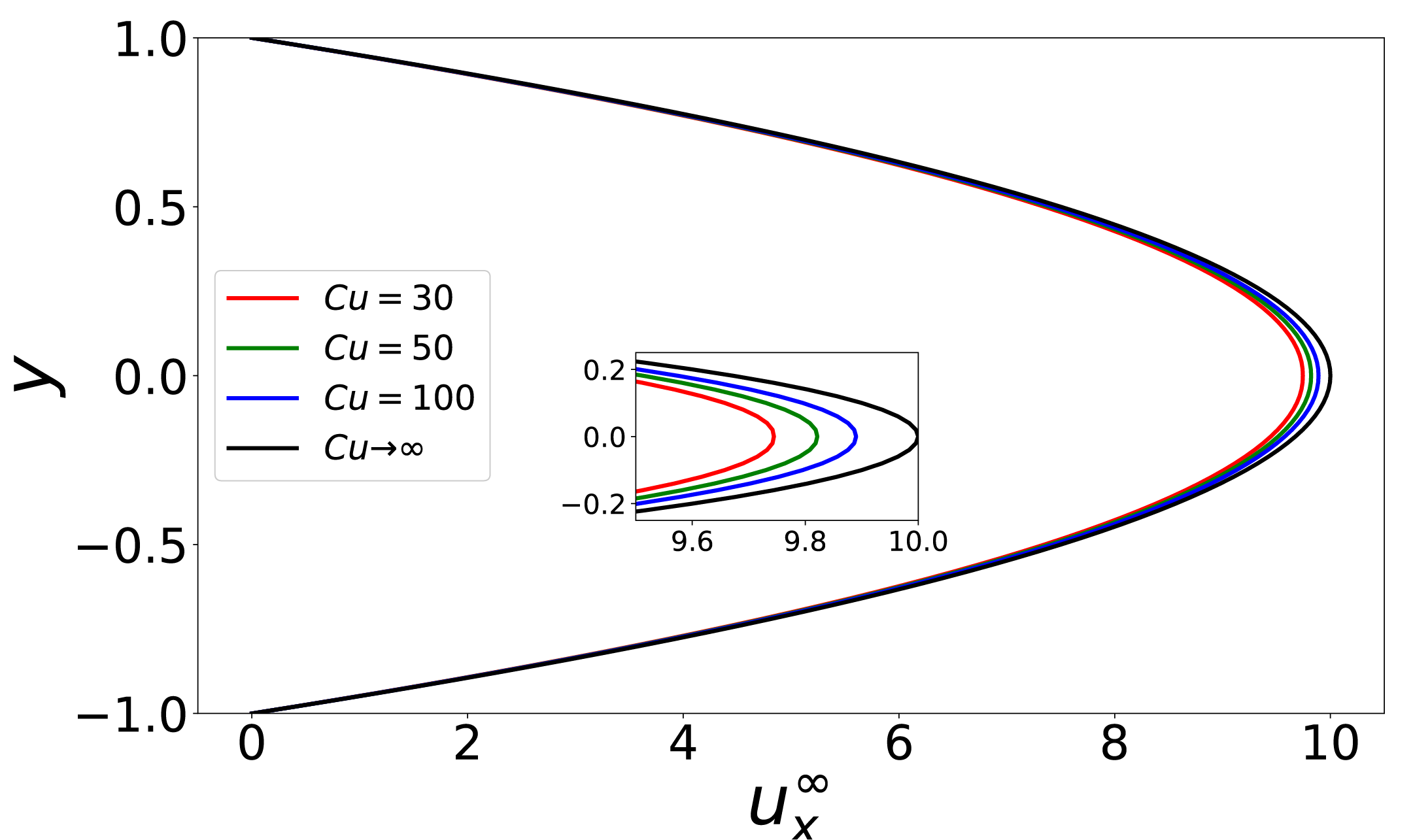}}
\caption{Pressure driven flow of shear thinning fluids in limits of (a) $Cu \ll 1$ (Eq.~\eqref{eq:background_low_Cu}) and (b) $Cu \gg 1$ (Eq.~\eqref{eq:background_high_Cu}).  The rheological parameters are $\epsilon =0.9$ and $n=0.3$. } 
\label{fig:Poiseuille}
\end{figure}

Below state the results for the velocity field in terms of the non-dimensional quantities of the Carreau model (Table 1).  These results were obtained by performing a regular perturbation expansion in $Cu^2$ for small Carreau number, and $Cu^{n-1}$ for large Carreau number.  For small Carreau number ($Cu \ll 1$), the velocity field is:
\begin{subequations} \label{eq:background_low_Cu}
\begin{align} 
u_x^{\infty} = u_x^{\infty, (0)} + Cu^2 u_x^{\infty, (1)} + O(Cu^4)   \\
u_x^{\infty, (0)} = 1 – y^2; \qquad  u_x^{\infty, (1)} = \epsilon (1-n) (1 – y^4)
\end{align}
\end{subequations}
while for large Carreau number ($Cu \gg 1$) the velocity field is:
\begin{subequations}
\label{eq:background_high_Cu}
\begin{align}
u_x^{\infty} = u_x^{\infty, (0)} + Cu^{n-1} u_x^{\infty, (1)} + O(Cu^{2(n-1)}) \\
u_x^{\infty, (0)} = \frac{1 – y^2}{\beta}; \qquad  u_x^{\infty, (1)} = -\frac{2^n \epsilon}{\beta^{n+1}(n+1)} (1 – |y|^{n+1})
\end{align}
\end{subequations}
For both cases, the shear stress is $\tau_{xy} = -2y$.  Fig.~\ref{fig:Poiseuille} plots the flow field in the two perturbative limits.  The small Carreau approximation captures the deviation from the first Newtonian plateau, where the normalized viscosity $\mu =1$. A perturbation around that plateau gives rise to a reduced effective viscosity, ultimately resulting in a higher flow rate. Compare this with the large Carreau number approximation, which captures deviations around the second Newtonian plateau with a normalized viscosity $\mu =\beta.$ A perturbation around this plateau gives rise to a higher viscosity, which reduces the flow rate.

\subsection{Microhydrodynamics} \label{sec:microhydro}
When the spheroid’s size is much smaller than the channel height (i.e., $R \ll h$), one can determine the motion of the spheroid by assuming it is in an unbound fluid with a background velocity given by  Eqs.~\eqref{eq:background_low_Cu} and \eqref{eq:background_high_Cu}.  Here, we will use the coordinate system in Figure \ref{fig:Schematic_1}(b) to compute the particle's rigid body motion.  The coordinate $x_i$ represents the position from the particle's center of mass, while $u_i^{\infty}$ is the background flow from the previous section, transformed to the particle coordinate system.  We will use the reciprocal theorem to approximate expressions for the rigid body motion in the limit of small and large Carreau numbers.

\subsubsection{Problem setup, reciprocal theorem, and perturbation expansion}

Suppose we have an inertialess, incompressible fluid that satisfies the Carreau constitutive relationship Eqs.~\eqref{eq:Carreau}.  This stress tensor can be decomposed into two components:

\begin{subequations} \label{eqn:stress_decomposition}
\begin{align}
\tau_{ij} &= \tau_{ij}^N + \tau_{ij}^{ex} \\
\tau_{ij}^N &= \mu \dot{\gamma}_{ij} – p \delta_{ij} \\
\tau_{ij}^{ex} &= (\eta - \mu) \dot{\gamma}_{ij}
\end{align}
\end{subequations}
where $\tau_{ij}^N$ is a Newtonian stress tensor with constant viscosity $\mu$, while $\tau_{ij}^{ex}$ is an extra stress tensor that depends on the shear-thinning viscosity $\eta$ described in Eq.~\eqref{eq:consti_dimless}.  

When we place a spheroid in such a fluid with far field velocity $u_i^{\infty}$, the continuity and momentum equations outside the particle can be written as:

\begin{equation} \label{eq:mom_balance}
    \frac{\partial u_i}{\partial x_i} = 0; \qquad \frac{\partial \tau_{ij}^N}{\partial x_j} + b_i = 0
\end{equation}
where 
\begin{equation}
    b_i = \frac{\partial}{\partial x_j} \left( \tau_{ij}^{ex} \right)
\end{equation}
is an effective body force on a Newtonian fluid.  Far away from the particle ($|x_i| \rightarrow \infty$), the velocity and stress fields are $u_i^{\infty}$, $\tau_{ij}^{N,\infty}$, and $\tau_{ij}^{ex,\infty}$, which also satisfy the above relationship \eqref{eq:mom_balance} throughout the entire domain.  On the surface $S_p$ of the particle, the velocity is rigid body motion:

\begin{equation}
    u_i = U_i + \epsilon_{ijk} \Omega_j x_k \qquad \text{on } \qquad x_i \in S_p
\end{equation}
where $U_i$ and $\Omega_i$ are the translational and rotational velocities of the particle.  The force and torque balance also need to be satisfied.  The external force and torque on the particle are:

\begin{subequations}
\begin{align}
    F_i^{ext} &= \int_{S_p} \tau_{ij} n_j dS \\
    T_i^{ext} &= \int_{S_p} \epsilon_{ijk} x_j \tau_{km} n_m dS
\end{align}
\end{subequations}
where the normal vector $n_i$ points inward to the particle.  In this problem, $F_i^{ext} = T_i^{ext} = 0$.  The force and torque on the particle are also zero from the far-field stress $\tau_{ij}^{\infty}= \tau_{ij}^{N,\infty} + \tau_{ij}^{ex,\infty}$.

We note that the above problem is formulated as the Stokes flow around a particle with a far field velocity $u_i^{\infty}$ and a fluid body force $b_i$.  In this situation, one can employ the reciprical theorem to obtain an expression for the rigid body motion $(U_i, \Omega_i)$.  This theorem has a celebrated history in the Stokes flow community -- see classical texts \cite{KimKarilla2005,HB83}.  Below, we provide the final formula for $(U_i, \Omega_i)$ -- a detailed derivation can be found in Appendix A.

The expression for the particle's rigid body motion satisfies the following relationship:

\begin{equation} \label{eq:reciprocal_final}
    \begin{bmatrix}
R^{FU}_{ij} & R^{F\Omega}_{ij}\\
R^{TU}_{ij} & R^{T\Omega}_{ij}
\end{bmatrix}     \begin{bmatrix}
U_j \\
\Omega_j 
\end{bmatrix} = \begin{bmatrix}
F_i^{eff} \\
T_i^{eff} 
\end{bmatrix}
\end{equation}

In the above equation, the quantities  $R_{ij}^{FU}$, $R_{ij}^{F\Omega}$, $R_{ij}^{TU}$ and $R_{ij}^{T\Omega}$ are the Stokes-flow resistance tensors for the particle with Newtonian viscosity $\mu$.  This formula states that the the translation and rotation of the particle is equal to its motion in Stokes flow with an effective force and torque:

\begin{subequations} \label{eq:effective_force_torque}
\begin{align}
    F_i^{eff} = F_i^{ext} + F_i^{flow} + F_i^{poly}\\
    T_i^{eff} = T_i^{ext} + T_i^{flow} + T_i^{poly}
\end{align}
\end{subequations}

The first terms in the above expression correspond to the external force and torque ($F_i^{ext}$, $T_i^{ext}$) on the particle, which are zero for this problem.  The second terms correspond to the effective force and torque from the external flow $u_i^{\infty}$.  These quantities are:

\begin{subequations}  \label{eq:flow_force}
\begin{align}
    F_k^{flow} &= \int_{S_p} u_i^{\infty} \Sigma_{ijk}^{trans} n_j dS\\
    T_k^{flow} &= \int_{S_p} u_i^{\infty} \Sigma_{ijk}^{rot} n_j dS 
\end{align}
\end{subequations}

In the above formula, the quantities $\Sigma_{ijk}^{trans}$ and $\Sigma_{ijk}^{rot}$ are the the $ij$ components of the stress field on the particle surface in Stokes flow due to unit translation or unit rotation in the $k$ direction.  The normal vector $n_j$ points into the particle, and the expression is integrated over the particle surface $S_p$. 

The last terms in Eq.~\eqref{eq:effective_force_torque} are the effective force and torque from the non-Newtonian stress.  These quantities are:

\begin{subequations} \label{eq:poly_force}
\begin{align}
    F_k^{poly} &= - \int_{V} \frac{\partial v_{ik}^{trans}}{\partial x_j} \left( \tau_{ij}^{ex} - \tau_{ij}^{ex,\infty} \right) dV\\
    T_k^{poly} &= - \int_{V} \frac{\partial v_{ik}^{rot}}{\partial x_j} \left( \tau_{ij}^{ex} - \tau_{ij}^{ex,\infty} \right) dV 
\end{align}
\end{subequations}
where the volume of integration is outside the particle.  The quantities $v_{ik}^{trans}$ and $v_{ik}^{rot}$ represent the Stokes velocity field outside the particle in the $i$ direction due to unit translation or unit rotation in the $k$ direction. The quantity $\tau_{ij}^{ex}$ is the extra stress outside the particle, while $\tau_{ij}^{ex,\infty}$ is the extra stress in the absence of the particle.

We will now perform a perturbation expansion to estimate the rigid body motion $(U_i, \Omega_i)$ in the limits of small and large Carreau numbers.  We will expand the velocity fields as follows:
\begin{subequations}
\begin{align}
u_i &= u_i^{(0)} + \delta u_i^{(1)} + O(\delta^2) \\
u_i^{\infty} &= u_i^{\infty, (0)} + \delta u_i^{\infty, (1)} + O(\delta^2)
\end{align}
\end{subequations}
and similarly expand the viscosity of the fluid as:

\begin{equation}
    \eta = \mu + \delta A[\dot{\gamma}^{(0)}] + O(\delta^2)
\end{equation}
In the above equations, $\delta$ is a small parameter, where $\delta = Cu^2$ for small Carreau number $(Cu \ll 1)$, and $\delta = Cu^{(n-1)}$ for large Carreau number $(Cu \gg 1)$.  For small Carreau number, the constant viscosity is $\mu = 1$, while the nonlinear function is $A[\dot{\gamma}] = \frac{1}{2} \epsilon (n - 1) \dot{\gamma}^2$.  This situation corresponds to a weak departure from the zero shear rate plateau.  For large Carreau number, the constant viscosity is $\mu = \beta$, while $A[\dot{\gamma}] = \epsilon \dot{\gamma}^{n-1}$.  This situation corresponds to a weak departure from the infinite shear rate plateau.

We will now plug in the above expansions into Eqs.~\eqref{eq:reciprocal_final}-\eqref{eq:poly_force} to obtain the particle's translation and rotational velocity $(U_i, \Omega_i)$ to $O(\delta)$.  We specifically require the far-field velocity $u_i^{\infty}$ up to $O(\delta)$ for the flow force and torque (Eq.~\eqref{eq:flow_force}).  The expression for $u_i^{\infty}$ is found in Eqs.~\eqref{eq:background_low_Cu} and \eqref{eq:background_high_Cu}, which needs to be transformed to the particle coordinate system.  We also require the extra stress tensor $\tau_{ij}^{ex} - \tau_{ij}^{ex,\infty}$ up to $O(\delta)$ for the polymeric force and torque (Eq.~\eqref{eq:poly_force}).  The expression for this quantity is:

\begin{equation} \label{eq:extra_stress_integrand}
    \tau_{ij}^{ex} - \tau_{ij}^{ex, \infty} = \delta A[\dot{\gamma}^{(0)}] \dot{\gamma}_{ij}^{(0)} - \delta A[\dot{\gamma}^{\infty, (0)}] \dot{\gamma}_{ij}^{\infty, (0)} + O(\delta^2)
\end{equation}
where as mentioned before,  $A[\dot{\gamma}] = \frac{1}{2} \epsilon (n - 1) \dot{\gamma}^2$ in the small Carreau number limit ($Cu \ll 1)$, while $A[\dot{\gamma}] = \epsilon \dot{\gamma}^{n-1}$ in the large Carreau number limit $(Cu \gg 1)$.  The above expression depends on the $O(1)$ shear rate tensors $\dot{\gamma}^{(0), \infty}_{ij}$ and $\dot{\gamma}^{(0)}_{ij}$, which are spatial derivatives of the $O(1)$ velocity fields.  The $O(1)$ velocity field $u_i^{(0)}$ is the Stokes flow around the particle with a far-field velocity $u_i^{\infty, (0)}$.  For the specific case where the particle is an ellipsoid and the far-field velocity is parabolic (e.g., Eqs.~\eqref{eq:background_low_Cu} and \eqref{eq:background_high_Cu} at $O(1)$), the solution to $u_i^{(0)}$ is known \cite{Martin2019SurfaceFields,Wang2019ImproperEllipsoid}.

\subsubsection{Formulae for ellipsoids}

Eqs.~\eqref{eq:reciprocal_final}-\eqref{eq:poly_force} describe the rigid body motion of a particle in a non-Newtonian fluid.  For a specified velocity field $u_i^{\infty}$ and extra stress tensor $\tau_{ij}^{ex}$, one can compute the particle's translational and rotational velocity $(U_i, \Omega_i)$.  Performing this calculation requires knowledge of the particle's Stokes flow behavior.  Specifically, one needs to know the velocity fields $(v_{ik}^{trans}, v_{ik}^{rot})$ and stress fields $(\Sigma_{ijk}^{trans}, \Sigma_{ijk}^{rot})$ outside the particle arising from rigid body motion, and one needs to know the resistance tensors ($R_{ij}^{FU}$, $R_{ij}^{F\Omega}$, $R_{ij}^{TU}, R_{ij}^{T\Omega}$).  These quantities are well-known for an ellipsoid.



Let us consider an ellipsoid with semi-axes $(a,b,c) = (a_1, a_2, a_3)$, and let us choose a particle coordinate system that aligns with these axes.  In this situation, Eqs.~\eqref{eq:reciprocal_final}-\eqref{eq:poly_force} simplify considerably for the particle's rigid body motion.  For the case when the external force and torque are zero ($F_i^{ext}=T_i^{ext} = 0$), we obtain:

\begin{subequations} \label{eq:rigid_body_ellipse}
\begin{align}
    U_i &= U_i^{Faxen} + \frac{1}{R_{ii}^{FU}} F_i^{poly} \qquad \text{(no summation over i)}\\
    \Omega_i &= \Omega_i^{Faxen} + \frac{1}{R_{ii}^{T\Omega}} T_i^{poly} \qquad \text{(no summation over i)}
\end{align}
\end{subequations}
where $U_i^{Faxen}$ and $\Omega_i^{Faxen}$ are the translational and rotational velocities from Faxen's formula for an ellipsoid.  These quantities are:

\begin{subequations} \label{eq:Faxen_general}
\begin{align}
    U_i^{Faxen} = -\frac{1}{3V_p} \int_{S_p} u_i^{\infty} (n_j x_j) dS \\
    \Omega_i^{Faxen} = -\frac{1}{V_p} P_{ij} \epsilon_{jrk} \int_{S_p} x_r u_k^{\infty} (n_m x_m) dS
\end{align}
\end{subequations}
where $V_p = \frac{4\pi}{3} abc$ is the particle volume and $P_{ij}$ is a diagonal tensor.  The 11 component of this tensor is $P_{11} = \frac{1}{a_2^2 + a_3^2}$ with the other diagonal components obtained from index cyling.  If we Taylor expand the velocity field around the center of mass $(x_i = 0)$, one can get an approximate expression for the Faxen's velocities up to $O((R/h)^2)$, which is within the current current approximation of our model.  This procedure yields:

\begin{equation} \label{eq:Faxen_trans}
     U_i^{Faxen} = u_i^{\infty}(0) + \frac{1}{6} \sum_{k=1}^3 a_k^2 \frac{\partial u_i^{\infty}}{\partial x_k \partial x_k}(0) + \dots
\end{equation}
where the far field velocity $u_i^{\infty}$ and its derivatives are evaluated at the center of mass.  For the rotational velocity, one obtains:

\begin{equation} \label{eq:Faxen_rot}
    \Omega_1^{Faxen} = \frac{1}{2} \omega_1^{\infty}(0) + \frac{a_2^2 - a_3^2}{a_2^2 + a_3^2} E_{23}^{\infty}(0) + \dots
\end{equation}
where $\omega_i^{\infty} = \epsilon_{ijk}\frac{\partial u_k^{\infty}}{\partial x_j}$ is the vorticity evaluated at the center of mass, and $E_{ij}^{\infty} = \frac{1}{2} \dot{\gamma}_{ij}^{\infty}$ is the rate of strain tensor evaluated at the center of mass.  The other components for the rotational velocity are obtained by index cycling.


In Eq.~\eqref{eq:rigid_body_ellipse}, the resistance tensors $R_{ii}^{FU}$ and $R_{ii}^{T\Omega}$ take the following form:

\begin{subequations} \label{eq:resistance_tensors}
\begin{align}
    R_{11}^{FU} = 16\pi \mu abc\frac{1}{\chi_0 + \alpha_1 a_1^2}\\    R_{11}^{T\Omega} = \frac{16\pi \mu abc}{3} \frac{a_2^2 + a_3^2}{a_2^2 \alpha_2 + a_3^2 \alpha_3}
\end{align}
\end{subequations}

In the above equations, $\chi_0$ and $(\alpha_1, \alpha_2, \alpha_3)$ are elliptic integrals given in publication \cite{Wang2019ImproperEllipsoid}.  If one wants to obtain the other components of the diagonal tensor, one changes the indices for $a_i$ and $\alpha_i$ appropriately.  For example, for for $R_{22}^{FU}$, one changes the index to ``2'' while for $R_{22}^{T\Omega}$ one changes the indices to ``1'' and ``3''.



To evaluate the polymeric force $F_i^{poly}$ and torque $T_i^{poly}$ on the particle, one uses Eq.~\eqref{eq:poly_force}, which uses the Stokes velocity fields $v_{ik}^{trans}$ and $v_{ik}^{rot}$ outside the particle due to unit translation and unit rotation. Full expressions for $v_{ik}^{trans}$ and $v_{ik}^{rot}$ are given in Appendix B.  This formula also requires knowledge of the extra stress tensor $\tau_{ij}^{ex} - \tau_{ij}^{ex, \infty}$ (Eq. \eqref{eq:extra_stress_integrand}), which involves the $O(1)$ velocity field around the particle.  The solution to the $O(1)$ velocity field around an ellipsoid is calculated in papers \cite{Martin2019SurfaceFields,Wang2019ImproperEllipsoid} (see also \cite{Wang_Narsimhan_POF}).   We will refer readers to these works for the full expressions.











\section{Code development and verification}
\label{sec:code}

Given a particle’s initial position $(x_0, y_0, z_0)$, and orientation ($\theta_0, \phi_0$) in the channel (Fig. \ref{fig:Schematic_1} a), we track its position and orientation over a time as follows:

\begin{enumerate}
    \item \underline{Transform from lab coordinates to particle coordinates:}  We transform from lab coordinates (Fig.\ref{fig:Schematic_1} (a)) to particle coordinates (Fig.~\ref{fig:Schematic_1} (b)), where the origin is now at the spheroid’s center of mass and the axes are aligned with the spheroid’s semi-axes.  We transform the channel velocity field $u_i^{\infty}$ (Eqs.~\eqref{eq:background_low_Cu} and \eqref{eq:background_high_Cu}) to this coordinate system as well.  This velocity field is evaluated up to $O(\delta)$, where $\delta = Cu^2$ for the small Carreau number limit and $\delta = Cu^{n-1}$ for the large Carreau number limit.
    
    \item \underline{Evaluate Faxen velocities:}  Using the expression for $u_i^{\infty}$, we evaluate the Faxen velocities $U_i^{Faxen}$ and $\Omega_i^{Faxen}$ using Eqns.~\eqref{eq:Faxen_trans} and \eqref{eq:Faxen_rot}.  
    
    \item \underline{Calculate polymer force and torque:}  We calculate the polymer force and torque on the spheroid using Eq.~\eqref{eq:poly_force}.  For each component of the polymer force and torque, we numerically evaluate the volume integral in ellipsoidal coordinates using Gaussian quadrature.  We typically use $15$ quadrature points in the radial direction, $10$ quadrature points in the latitude direction, and $35$ quadrature points in the longitude direction. In the integrand, we use the expression for the extra stress $\tau_{ij}^{ex} - \tau_{ij}^{ex,\infty}$ described in Eq.~\eqref{eq:extra_stress_integrand}. Note that the integrand also involves the quantities from the auxiliary problem $ (v_{ik}^{trans}, v_{ik}^{rot})$, which has been solved and tabulated separately beforehand. 
    
    \item \underline{Calculate rigid body motion:}  Using the Faxen velocities and the polymer force/torque, we calculate the rigid body motion of the particle in the particle coordinate frame (Eqs.~\eqref{eq:rigid_body_ellipse} and \eqref{eq:resistance_tensors}).
    
    \item \underline{Transform back into lab coordinates and update particle position and orientation}:  We transform the rigid body velocities $(U_i, \Omega_i)$ back to the lab coordinates.  We then update the particle’s position and orientation.  The equation governing the particle’s center of mass is:
\begin{equation} \label{eq:com_evolution}
\frac{d x_i^{cm}}{d t} = U_i
\end{equation}
The equation governing the time evolution of the projection vector is $\frac{d p_i}{dt} = \epsilon_{ijk} \Omega_j p_k$.  If written in terms of the orientation angles $(\theta, \phi)$ of the spheroid, these equations become:
\begin{equation} 
\label{eq:projector_evolution}
\begin{gathered}
\frac{d \theta}{dt} =\Omega_{y} \cos \phi-\Omega_{x} \sin \phi \\
\frac{d \phi}{dt} =\Omega_{z}-\Omega_{x} \cos \phi \cot \theta-\Omega_{y} \sin \phi \cot \theta,
\end{gathered}
\end{equation}
	We update the particle’s position and orientation angles using the above equations \eqref{eq:com_evolution} and \eqref{eq:projector_evolution}.  We perform forward Euler time-stepping, with the time step chosen between $10^{-3} < \Delta t  < 10^{-2}$ in dimensionless units.
	
	\item \underline{Repeat the above steps:}  Repeat steps 1-5 until we have simulated the particle motion over a sufficient period of time.

\end{enumerate}

\begin{figure}[t]
\centering
{\includegraphics[width=0.65\linewidth]{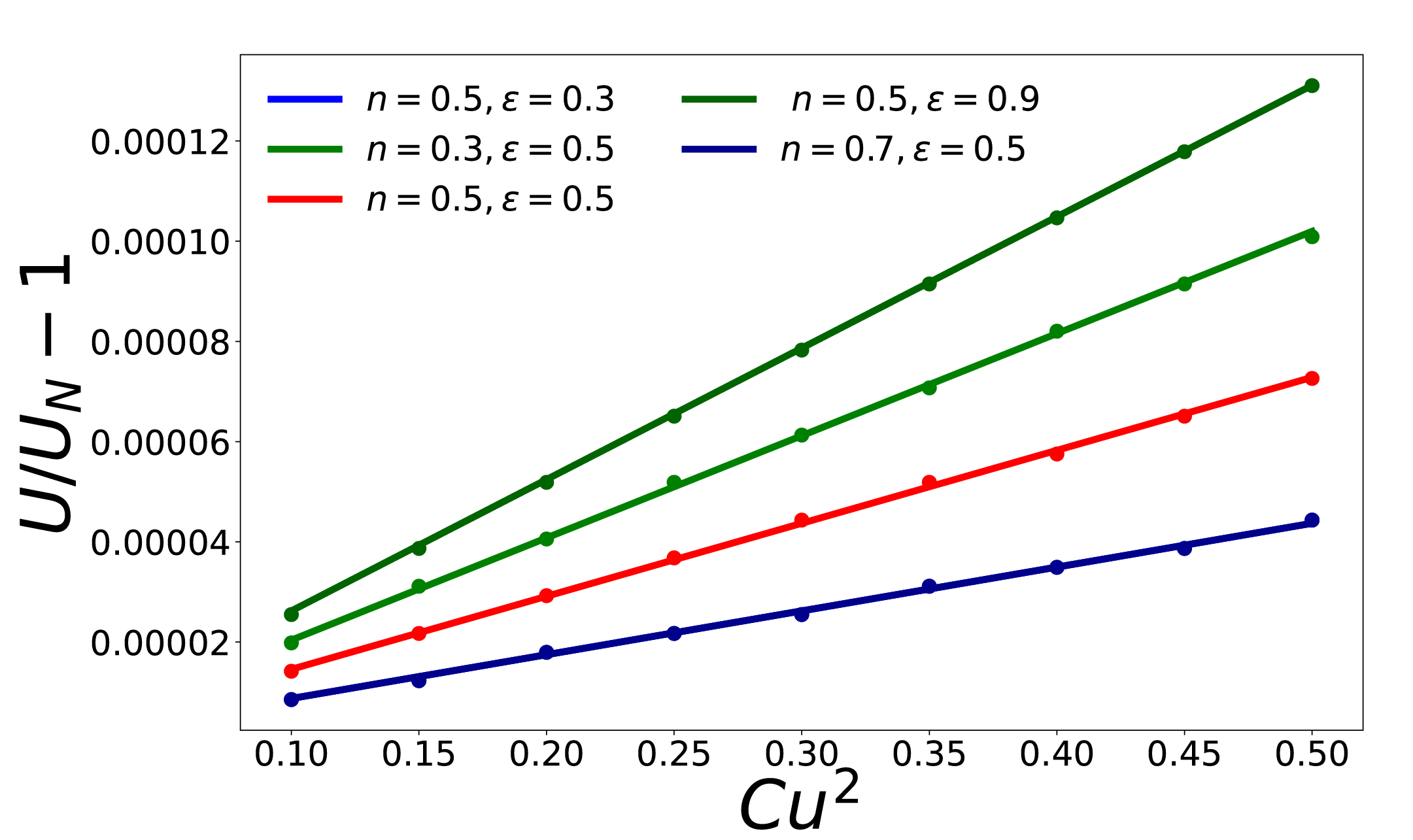}}
\caption{Sedimentation velocity of a sphere in a shear thinning fluid.  The $y$-axis shows the fractional increase of the sedimentation speed $U$ compared to the Newtonian value $U_N$ at zero Carreau number ($Cu = 0$).  The solid lines are results from the theory of \cite{Datta_Elfring_JNNFM_ShearThinning}, while the dots are the results of the reciprocal theorem based numerical simulation carried out in this paper.}
\label{fig:Validation}
\end{figure}

 We developed a MATLAB code to execute this algorithm. The code was initially developed and used for the analysis of an ellipsoid in a second order viscoelastic fluid in a previous publication \cite{Wang_Narsimhan_POF}. For this paper, we modified the code to account for shear thinning effects in both the low Carreau and high Carreau number limits. To test the code, we simulated the sedimentation of a falling sphere in a shear thinning fluid in the low Carreau number limit. Fig.~\ref{fig:Validation} show the results of our simulation and theoretical results provided by \cite{Datta_Elfring_JNNFM_ShearThinning}. The exact match between the results serves to verify our code.


\section{Results and discussion}
\label{sec:results_discussion}

In our discussion, we will stick to the orientational dynamics of spheroids and forego a discussion of translational motion.  We find that for translational motion, no novel phenomena are revealed like cross stream migration or particle lift \cite{Datta_Elfring_JNNFM_ShearThinning,Elfring_Main}.  We will first discuss the results in the small Carreau number limit, followed by the results in the large Carreau number limit.

Unless otherwise noted, we will examine a particle with equivalent particle radius $R = 0.1$ when rendered dimensionless by the channel height $h$. The initial position of the particle is at $[x_0,y_0,z_0]=[0,-0.5,0]$.




\subsection{Orientation dynamics -- small Carreau number limit ($Cu \ll 1$)}

\subsubsection{General observations – effect of $A_R$, $Cu$, and $n$:}

\begin{figure}[t]
\centering
\subfloat[]{\includegraphics[width=0.45\linewidth]{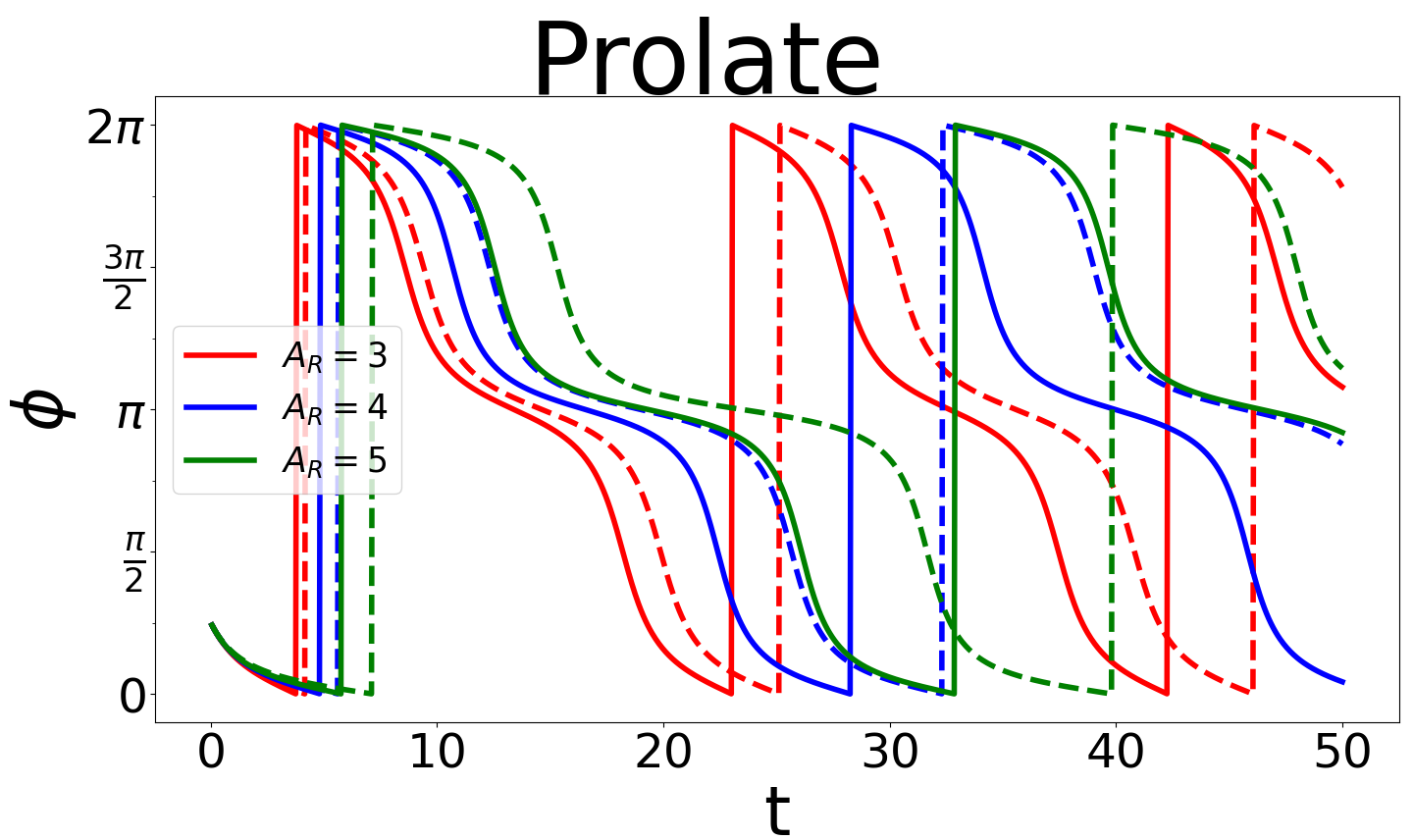}}
\subfloat[]{\includegraphics[width=0.45\linewidth]{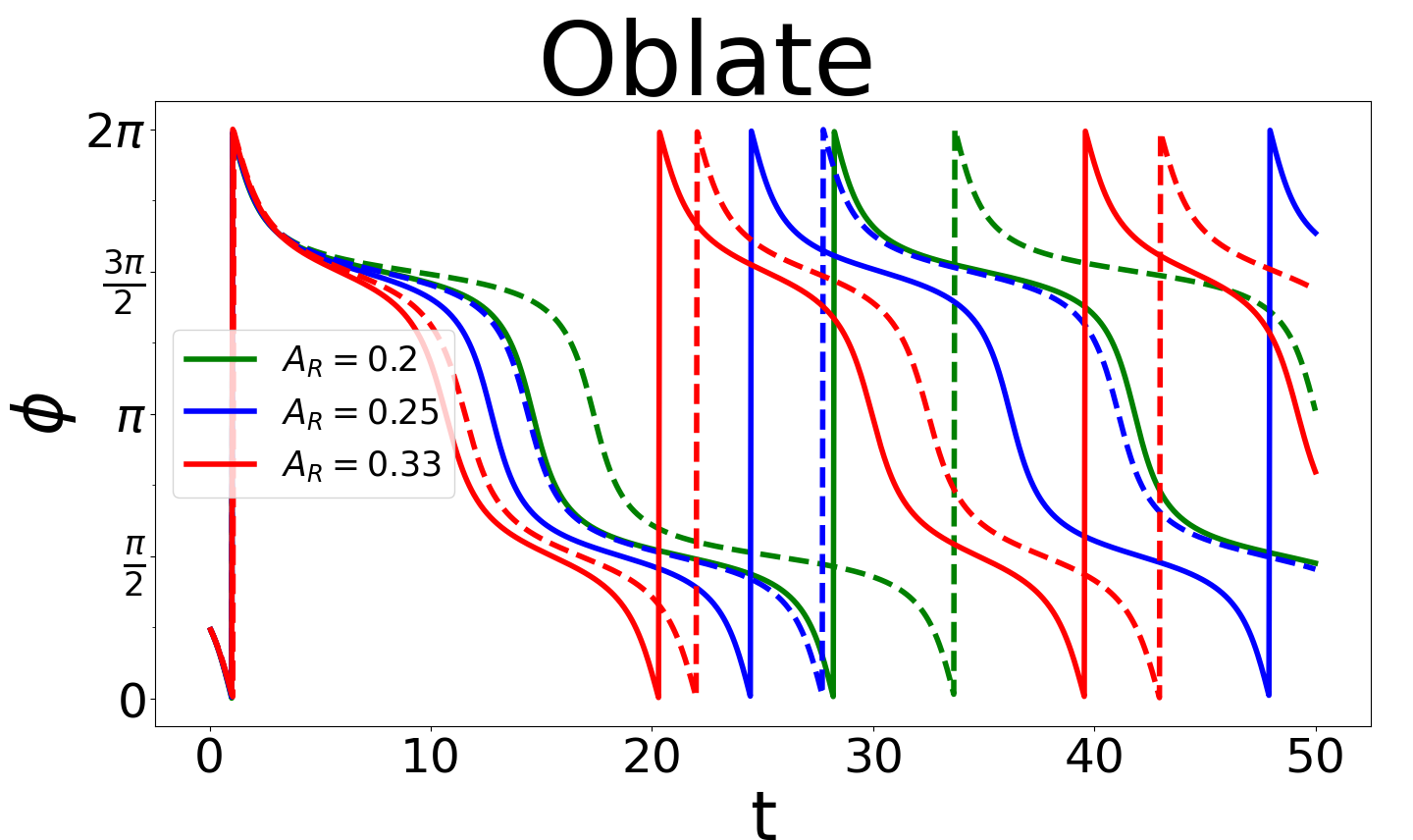}}
\hfill
\subfloat[]{\includegraphics[width=0.45\linewidth]{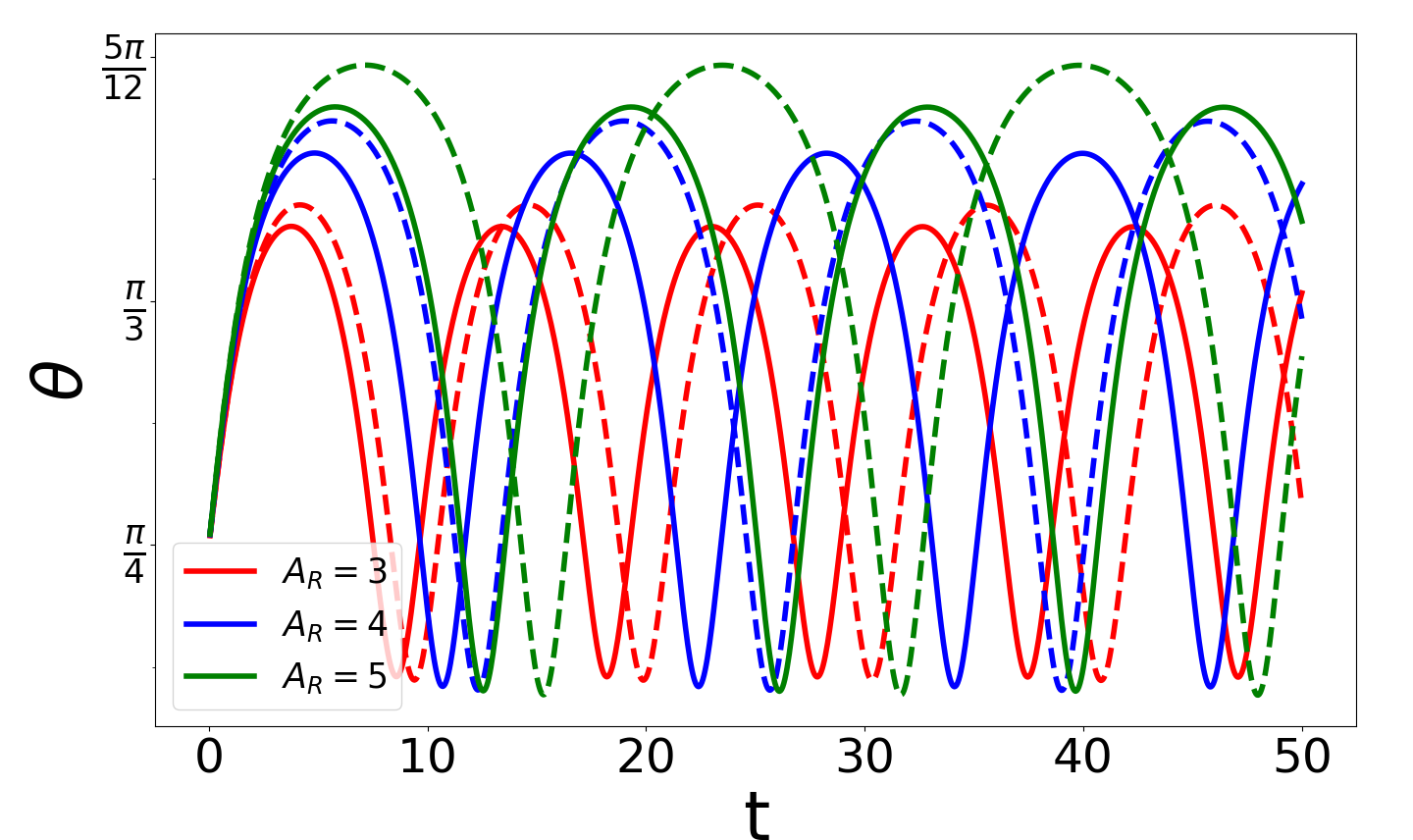}}
\subfloat[]{\includegraphics[width=0.45\linewidth]{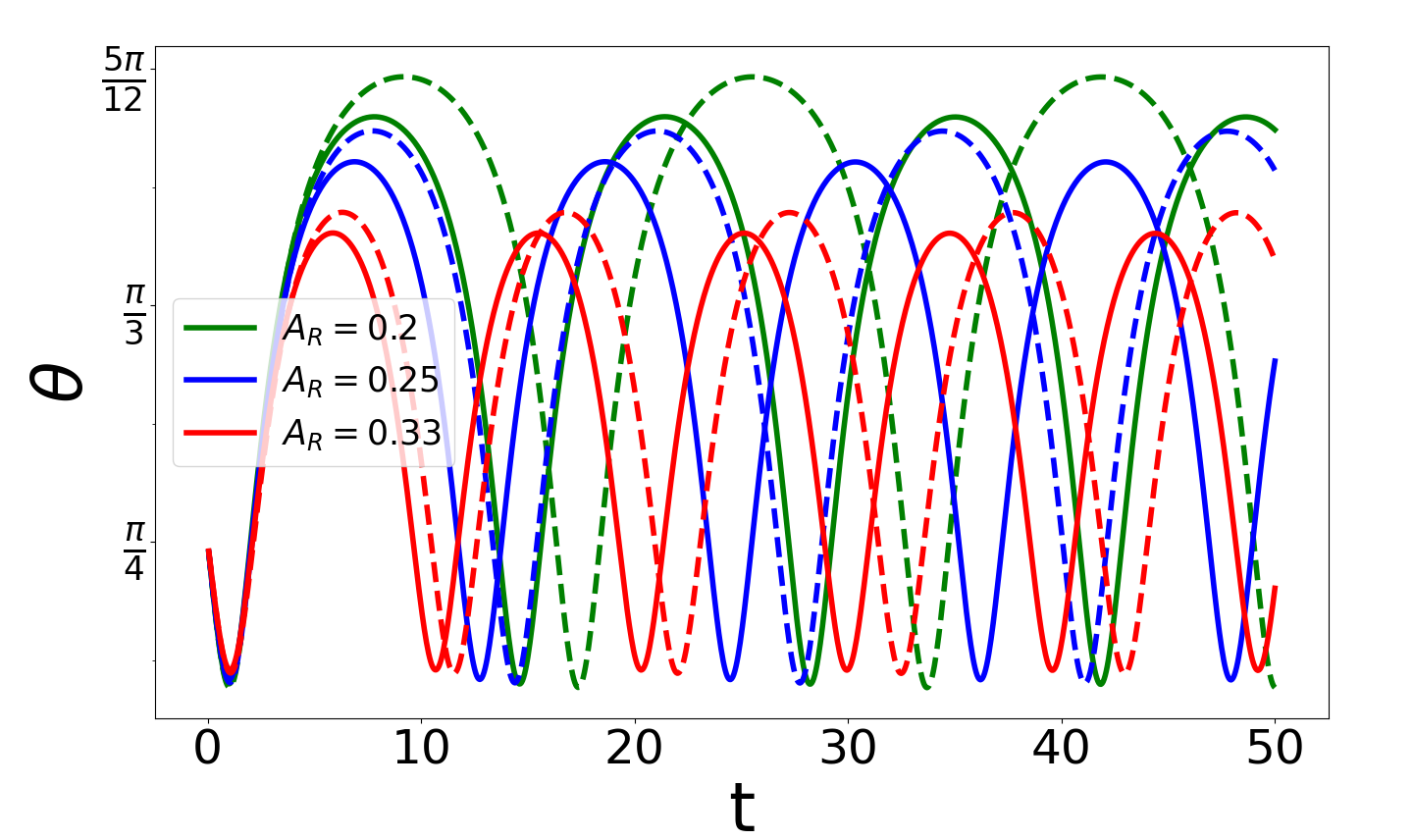}}
\hfill
\caption{Effect of shape parameter ($A_R$) on the orientation trajectories of prolate and oblate spheroids in the small Carreau number limit $(Cu \ll 1)$.  In all these plots, the initial orientation is $\theta_0 = \pi/4,\phi_0 =\pi/4$. The plots on the left show prolate spheroids ($A_R > 1$) and those on the right show oblate spheroids ($A_R < 1$). The solid curves denote shear thinning fluids with $Cu^2 =0.1$ and $n=0.3$, while the dashed curves denote Newtonian fluids ($Cu^2 = 0$).}
\label{fig:Low_AR}
\end{figure}



Fig.~\ref{fig:Low_AR} shows the influence of the shape parameter ($A_R$) on the orientational dynamics of spheroids in Newtonian and shear thinning fluids. First and foremost, we remark that the further a particle deviates from a spherical shape, the larger is the time period for its tumbling.  For a prolate particle ($A_R > 1$), the time period increases as $A_R$ increases.  For an oblate particle $(A_R < 1)$, the time period increases as $A_R$ decreases. These trends are clearly an artefact of Newtonian flow, where the time period scales as $T \sim (A_R+1/A_R)$. Next, we observe that the shear thinning reduces the tumbling time period compared to a Newtonian fluid —- see the solid (shear thinning) and dashed (Newtonian) curves in Fig.~\ref{fig:Low_AR}.  Additionally, shear thinning reduces the amplitude of oscillations in $\theta$ for the particles.  These trends are contrary to the trends observed for prolate spheroids in \textit{linear} flows of shear thinning fluids, as shown in the $Cu^2 \ll 1$ limit in the work of \cite{Elfring_Main}. In their case, shear thinning increases the time period of revolution for prolate spheroids, and increases the oscillation amplitude for $\theta$. This dichotomy in the trends is explained below.

For pressure driven flows analysed here, the deviatoric stress field balances the pressure gradient across the channel imposed on its ends. Since the imposed pressure drop is independent of the shear thinning correction, the stress induced in the flow does not change due to shear thinning. Since the stress field induced in the fluid is the same as that in the Newtonian case, a shear thinning fluid, being less viscous, has to deform more than a Newtonian fluid to sustain the same stress and pressure drop. Because the time period $T$ and the oscillation amplitude for $\theta$ scale as the inverse of strain rate/deformation, both these quantities decrease with an increase in shear thinning as shown in Fig.~\ref{fig:Low_AR}.

On the other hand, the linear (shear) flows studied by \cite{Elfring_Main} are kinematically controlled, wherein the velocity field of the background flow remains unchanged between Newtonian and shear thinning rheologies. Therefore, the stress field induced in a shear thinning flow field is weaker than that of a Newtonian field. A weaker stress field in turn imposes a diminished hydrodynamic torque on the particle, slowing down its rotation and increasing its oscillation amplitude $\theta$. The discussion of tumbling behavior of spheroids in shear flows and their comparison with the pressure driven flows is further elaborated upon in Sec. \ref{sec:LinearFlow}.

\begin{figure}[t]
\subfloat[]
{\includegraphics[width=0.5\linewidth]{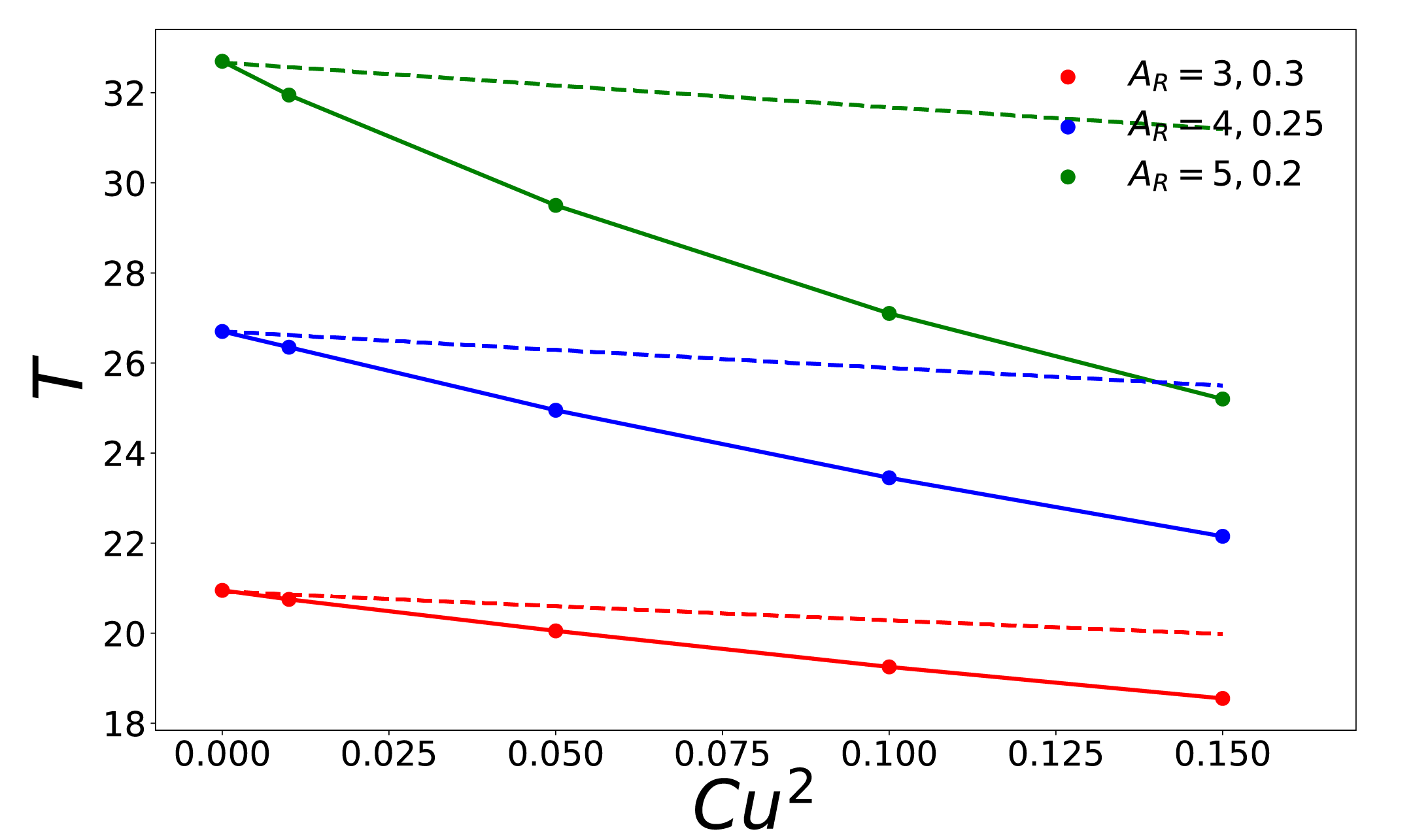}}
\subfloat[]
{\includegraphics[width=0.5\linewidth]
{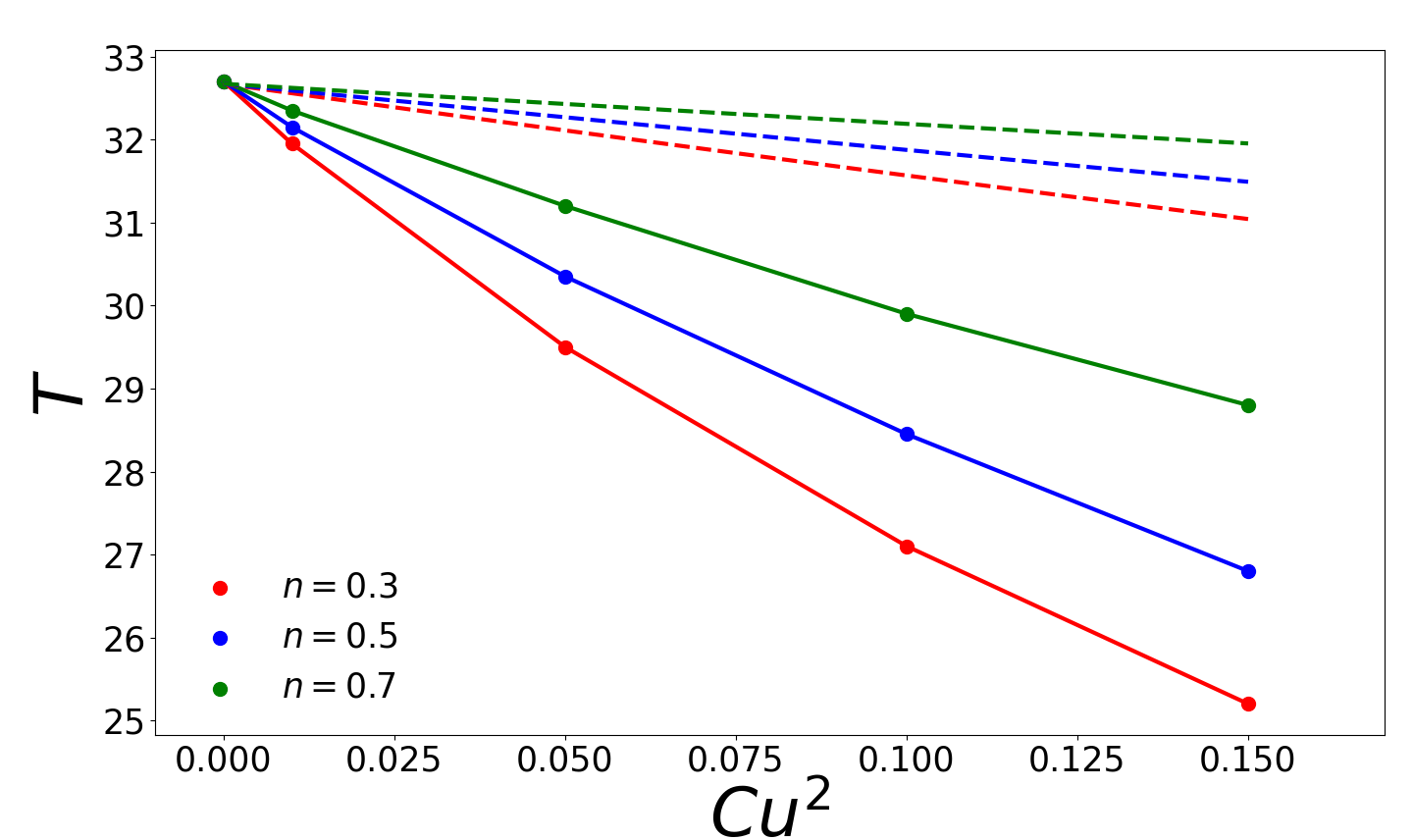}}
\caption{Tumbling time period for both prolate and oblate particles in the $Cu \ll 1$ limit for different values of (a) shape parameter $A_R$ and (b) power law index $n$. The initial orientation is $\theta_0 =\pi/4, \phi_0 =\pi/4$ for both the plots. The solid lines show the results from the reciprocal theorem based numerical estimation while the dotted lines show the result using Jeffrey's formula using the local shear rate -- i.e.,  $T =2\pi(A_R +1/A_R)\dot{\gamma}_{\text{loc}}^{-1}$, where $\dot{\gamma}_{\text{loc}} =\frac{\partial}{\partial y} \left[u_x^{\infty,(0)}+Cu^2 u_x^{\infty,(1)}\right]$}
\label{fig:Low_Time_AR_N}
\end{figure}

Fig.~\ref{fig:Low_Time_AR_N} plots the tumbling time period for prolate and oblate spheroids for different values of Carreau number ($Cu$), power law index ($n$) and shape parameter ($A_R$).  A couple of trends can be noticed here.  First, we clearly see that increased shear thinning (i.e., an increase in $Cu$ or a decrease in $n$) reduces the time period of tumbling, and this effect is amplified the further the particle deviates from a sphere. Even in the case of linear shear flows in \cite{Elfring_Main},  the effects of shear thinning are similarly enhanced at higher aspect ratios for prolate particles. The straightforward reason is that as the aspect ratio ($A_R$) of a prolate particle increases (for the same volume), the projection of the long axis along the velocity gradient $y$-axis (given by $a\sin{\theta}\sin{\phi}$) also becomes larger, and the particle is exposed more strongly to the shear thinning tendency of the flow. A similar argument holds for oblate particles as well. Interestingly, we see that time period is exactly equal when comparing the $A_R =3,4,5$ prolate particles to the $A_R =\frac{1}{3}, \frac{1}{4}, \frac{1}{5}$ oblate particles.  This observation suggests that the just like Newtonian case, the time period ${T}$ is  a function of $(A_R+1/A_R)$.



The last point we would like to illustrate is that the decrease in the tumbling time period cannot be wholly explained by the reduction in the effective shear rate around the particle.  In a Newtonian flow, the time period of tumbling is given by the classical Jeffrey formula $T =2\pi(A_R +1/A_R)\dot{\gamma}_{\text{loc}}^{-1}$, where the $\dot{\gamma}_{\text{loc}}$ is the local strain rate at the particle’s center of mass $y = y_0$.  A naïve approach to estimating the tumbling time period in a shear thinning fluid would be to use the same formula, noting that $\dot{\gamma}_{loc}$ increases in the channel as $Cu$ increases or $n$ decreases.  Fig.~\ref{fig:Low_Time_AR_N} shows the time period estimated using this approach, plotted as dotted lines.  Overall, while we see this approach accounts for some of the decrease in the particle’s time period, it does not match the simulations well.  This indicates that one cannot use simple modifications of Jeffrey’s theories to predict the time period of spheroids in shear thinning flows.


\subsubsection{Effect of initial orientation ($\theta_0, \phi_0$):}
\begin{figure}[t]
\centering
\subfloat[]{\includegraphics[width=0.45\linewidth]{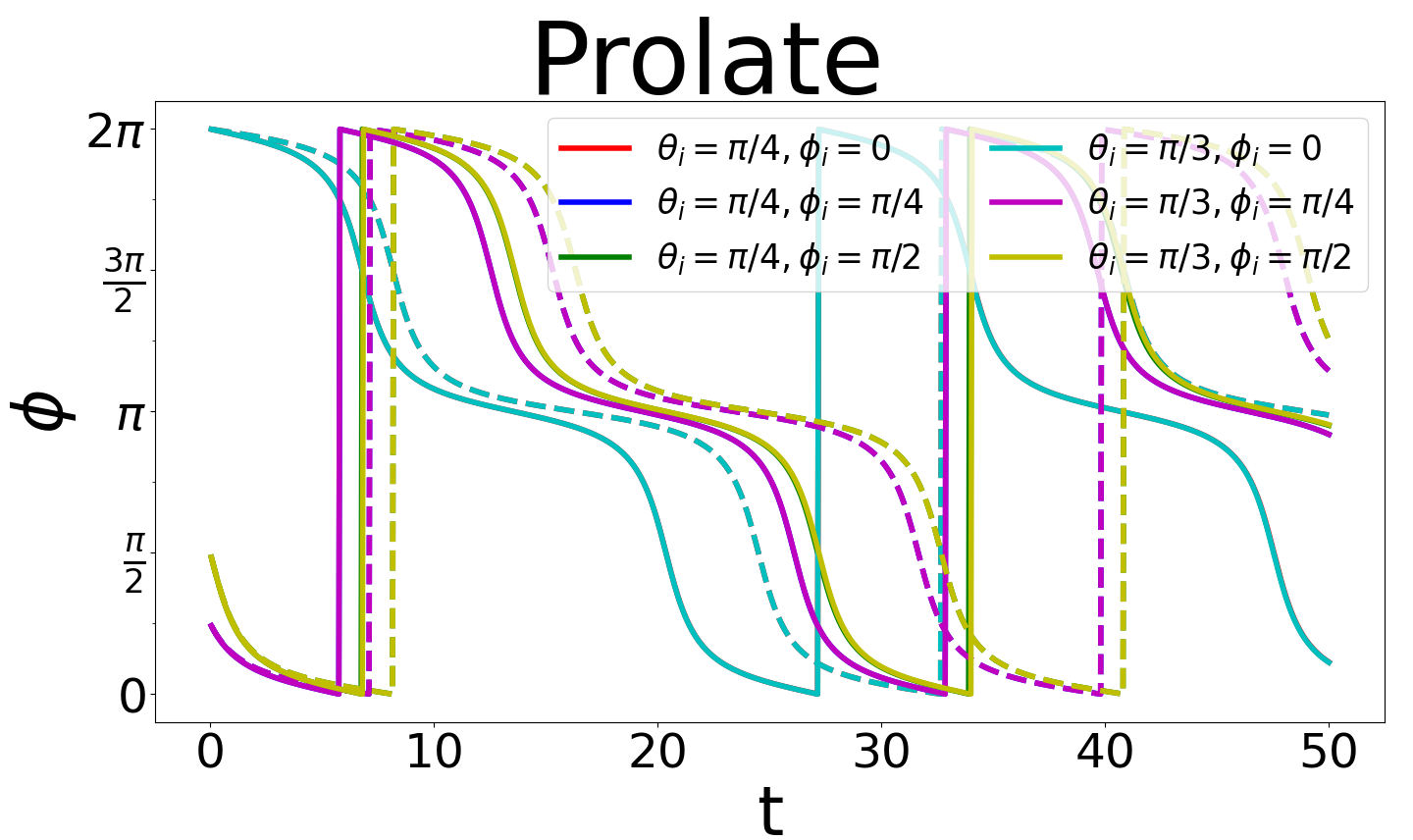}}
\subfloat[]{\includegraphics[width=0.45\linewidth]{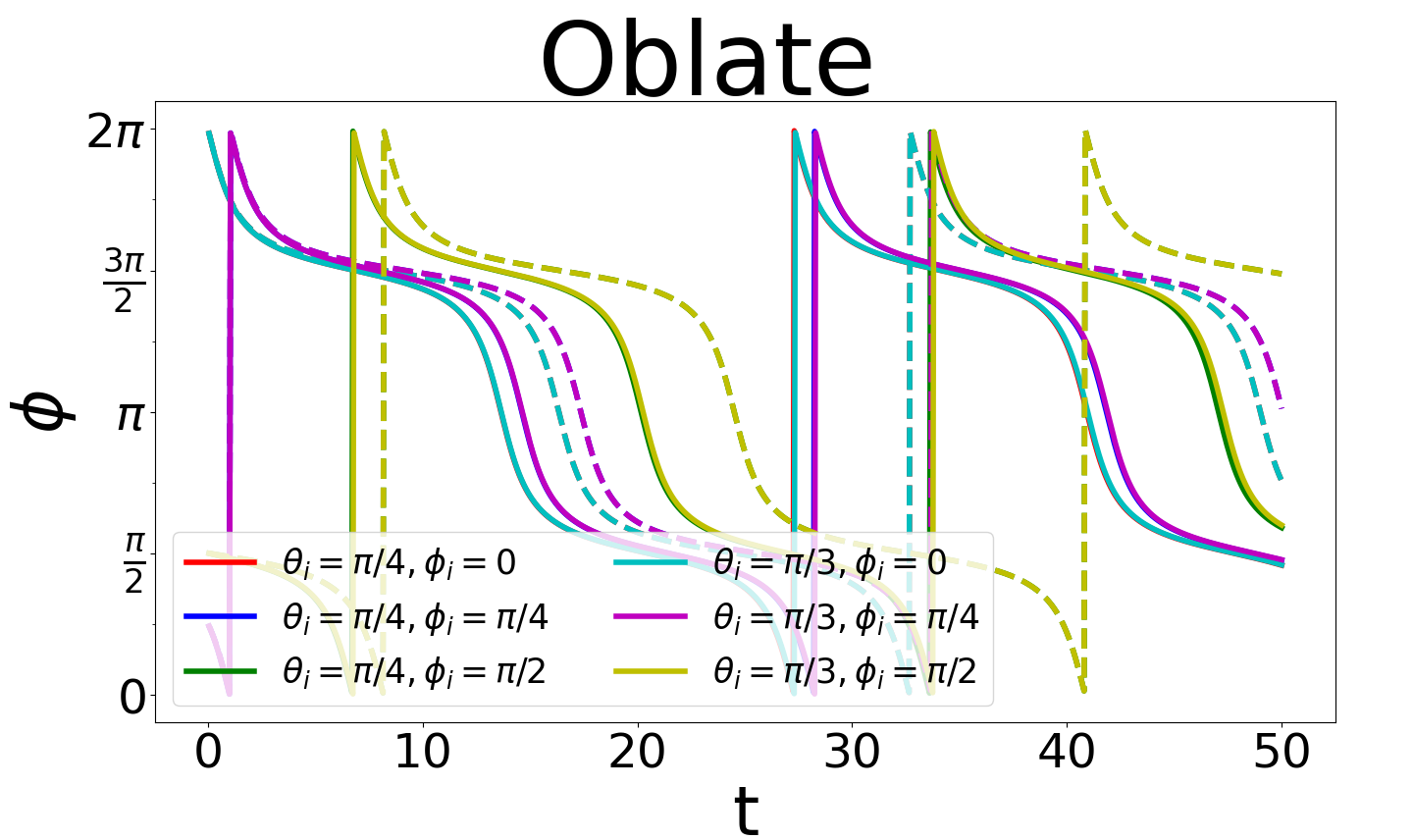}}
\hfill
\subfloat[]{\includegraphics[width=0.45\linewidth]{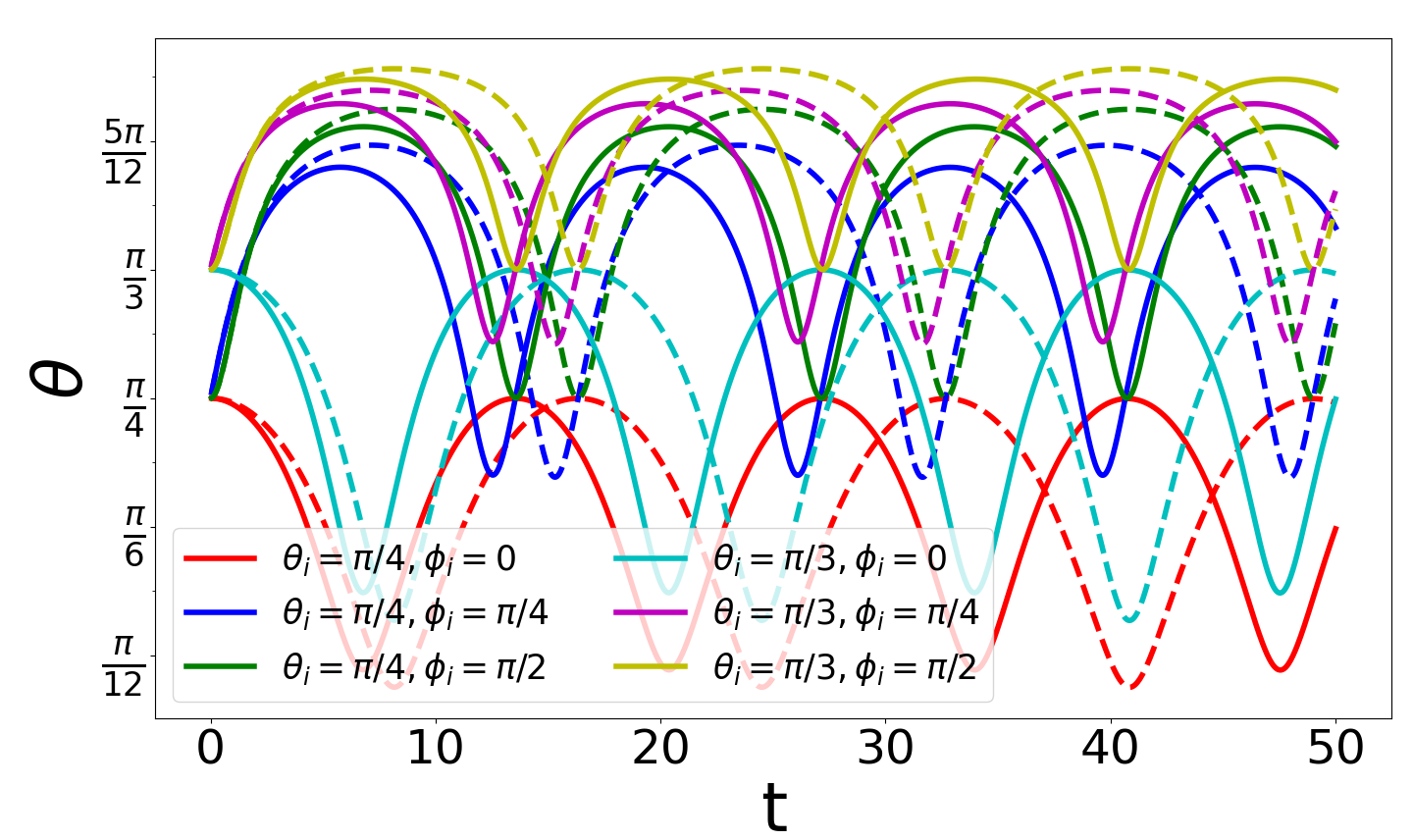}}
\subfloat[]{\includegraphics[width=0.45\linewidth]{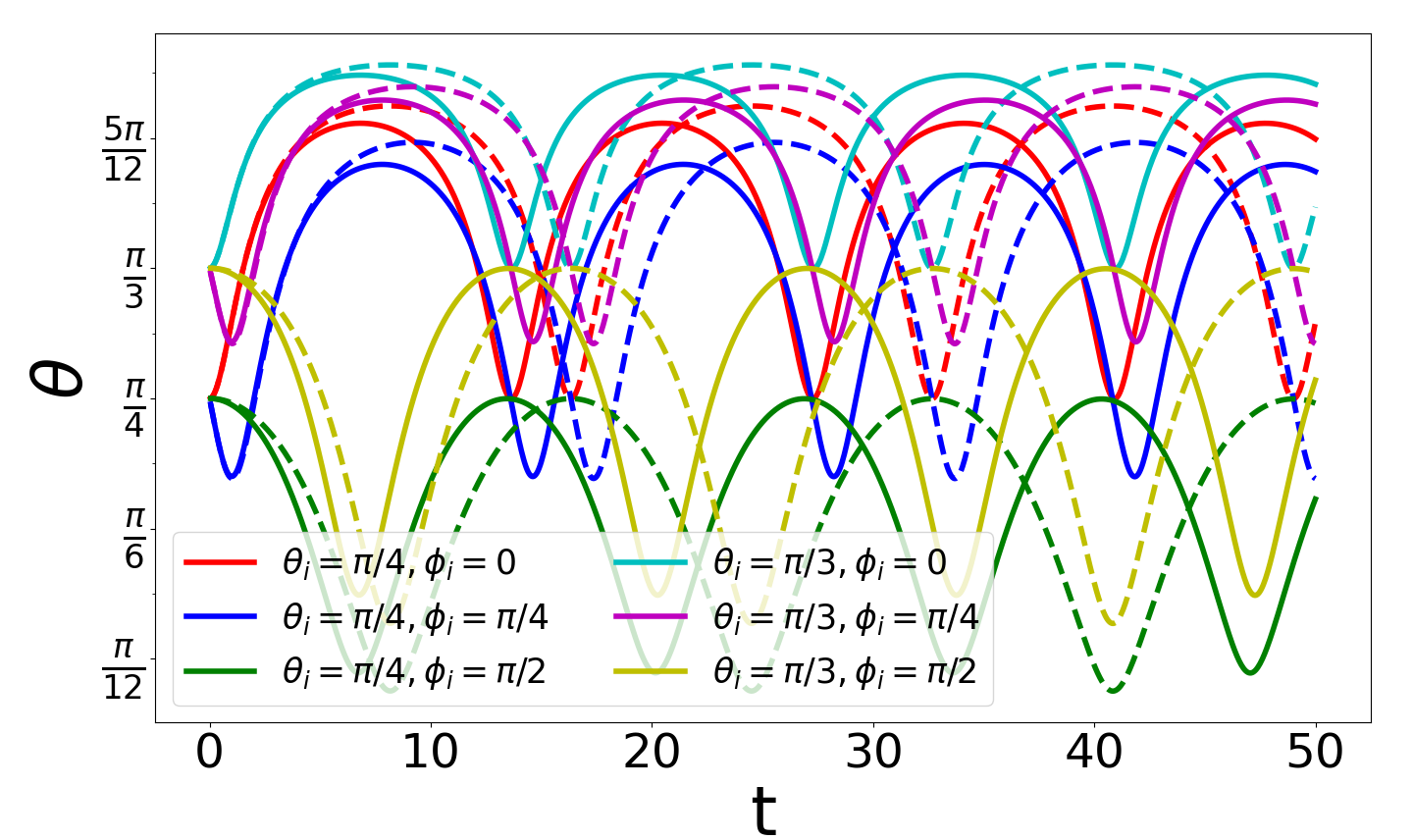}}
\caption{Effect of initial orientation ($\theta_0, \phi_0$) on the orientation trajectories of prolate and oblate spheroids in the small Carreau number limit $(Cu \ll 1)$.  The plots on the left show prolate spheroids with $A_R =5$ and those on the right show oblate spheroids with $A_R = 0.2$. The solid lines denote shear thinning flows with power law index $n =0.3$ and $Cu^2 =0.1$, while the dashed lines denote Newtonian fluids ($Cu^2 = 0$).}
\label{fig:Low_OR}
\end{figure}


Next, we analyse the effect of initial orientation ($\theta_0,\phi_0$) on the spheroid’s trajectory  in both Newtonian and shear thinning flows. There are six pairs of initial orientations we study: $(\theta_0,\phi_0) =$ ($(\pi/4,0)$, $(\pi/4,\pi/4)$, $(\pi/4,\pi/2)$, $(\pi/3,0)$, $(\pi/3,\pi/4)$, $(\pi/3,\pi/2)$). Both prolate and oblate particles are analysed.

First we observe from Fig.~\ref{fig:Low_OR} (a,b) that the evolution of the azimuth angle $\phi$ over time is independent of the initial orientation $\theta_0$ --in other words, curves for both $\theta_0 = \pi/4$ and $\theta_0 =\pi/3$ coincide for the same initial angle $\phi_0$.  For the prolate particle, the spheroid spends more time in $\phi =m \pi$ configuration ($m = 0,1$), while the oblate particle spends more time in the $\phi=(m+\frac{1}{2})\pi$ configuration ($m = 0,1$). The regions with $\phi = (m+\frac{1}{2})\pi$ (prolate particles) and $\phi =m\pi$ (oblate particles) show the largest deviation between Newtonian and shear thinning trajectories.  We can understand these trends by noting that the rotation of the spheroid is actuated by the hydrodynamic torque from the shear stresses. This torque arises from a traction in the $x$-direction and a moment arm (projection of particle length) along the $y$-axis. Thus, configurations with the smallest moment arm (e.g., $\phi =m\pi$ for prolate particle, $\phi =(m+\frac{1}{2})\pi$ for oblate particle) have the smallest rotation rate and longest residence time, while configurations with the longest moment arm (e.g., $\phi =(m+\frac{1}{2})\pi$ for prolate particle, and $\phi =m\pi$ for the oblate particle) have the largest rotation rate and hence largest influence of shear thinning.

At this juncture, we would like to compare the evolution of $\phi$ in this flow with spheroids in other kinds of flow. For prolate spheroids in pure shear flow of shear thinning fluids, the trend is same as observed here wherein the prolate spheroid rotates the fastest at $\phi =(m+\frac{1}{2})\pi$ and slowest in the $\phi =m\pi$ configuration (see Fig.4 (b) and Fig.4 (c) in the \cite{Elfring_Main}). On the other hand, for the case of spheroids in quadratic flow of nonlinear viscoelastic fluids, the reverse is true. For prolate (oblate) spheroids in such flows, the maxima (minima) of $\dot{\phi}$ occurs at $\phi =0$ or $\phi = \pi$, while the minima (maxima) occurs at $\phi =\pi/2$ (see  Fig. $6$ and the relevant discussion in Sec. IV c in \cite{Wang_Narsimhan_POF}). The reason behind this dichotomy is explained by considering the fact that in nonlinear viscoelastic flows, the hydrodynamic torque acting on the spheroid is induced by normal stresses, as opposed to shear stresses in shear thinning flows. These normal stresses give rise to a traction in the $y$-direction and a moment arm in the $x$-direction. Consequently, when the spheroid is oriented such that its longest axis is aligned with the $x$-axis, the hydrodynamic torque acting on the spheroid is the largest, resulting in higher rotation rate in this configuration.

We next analyse the evolution of $\theta$ in time for both Newtonian and shear thinning flows (Fig.~\ref{fig:Low_OR} (c,d)). From the classical treatment of a Jefferey orbit in a Newtonian fluid, we understand that $\dot{\theta}$ varies as $\frac{A_R^2-1}{A_R^2+1} \sin{2\theta}\sin{2\phi}$. We also observe similar features of this equation in shear thinning fluids. For example, we see $\dot{\theta}$ change between prolate and oblate particles of the same initial orientation due to the sign change in $\frac{A_R^2-1}{A_R^2+1}$ (compare the same color curves in the Fig.~\ref{fig:Low_OR} (c,d)).  We also see that a spheroid initially released at $\theta_{0} =\pi/2$ or $\theta_0 = 0$ will have $\dot{\theta}= 0$, and therefore remain at the initial angle.  Overall, we see shear thinning reduces the period and amplitude of $\theta$; however, the influence of the initial orientation on these quantities is seen to be less dominant than the other effects discussed earlier ($A_R$, $Cu$, and $n$) .  Therefore, we do not comment on this effect further in the manuscript.

\subsection{Orientation dynamics -- large Carreau number limit ($Cu \gg 1$)}

We now analyse the rotational dynamics of a spheroid in the large Carreau number limit ($Cu \gg 1)$, which captures deviations in the fluid viscosity from the lower Newtonian plateau $\eta = \beta$.
There are a couple of differences between the mathematical behavior in the   $Cu \ll 1$ limit and the $Cu \gg 1$ limit. First, since the $Cu \gg 1$ limit captures the deviation from the second Newtonian plateau, any perturbation from this plateau will increase the effective viscosity of the fluid. Secondly, for the large Carreau number limit, the second Newtonian plateau is recovered solely in the limit $Cu \to \infty$, but not in the limit $n \to 1$. This is in contrast to the small Carreau number limit where the upper pleateau is recovered by $n \to 1$  or $Cu \to 0$ or both.

\subsubsection{General observations – effect of $A_R$, $Cu$, and $n$}

\begin{figure}[t]
\centering
\subfloat[]{\includegraphics[width=0.45\linewidth]{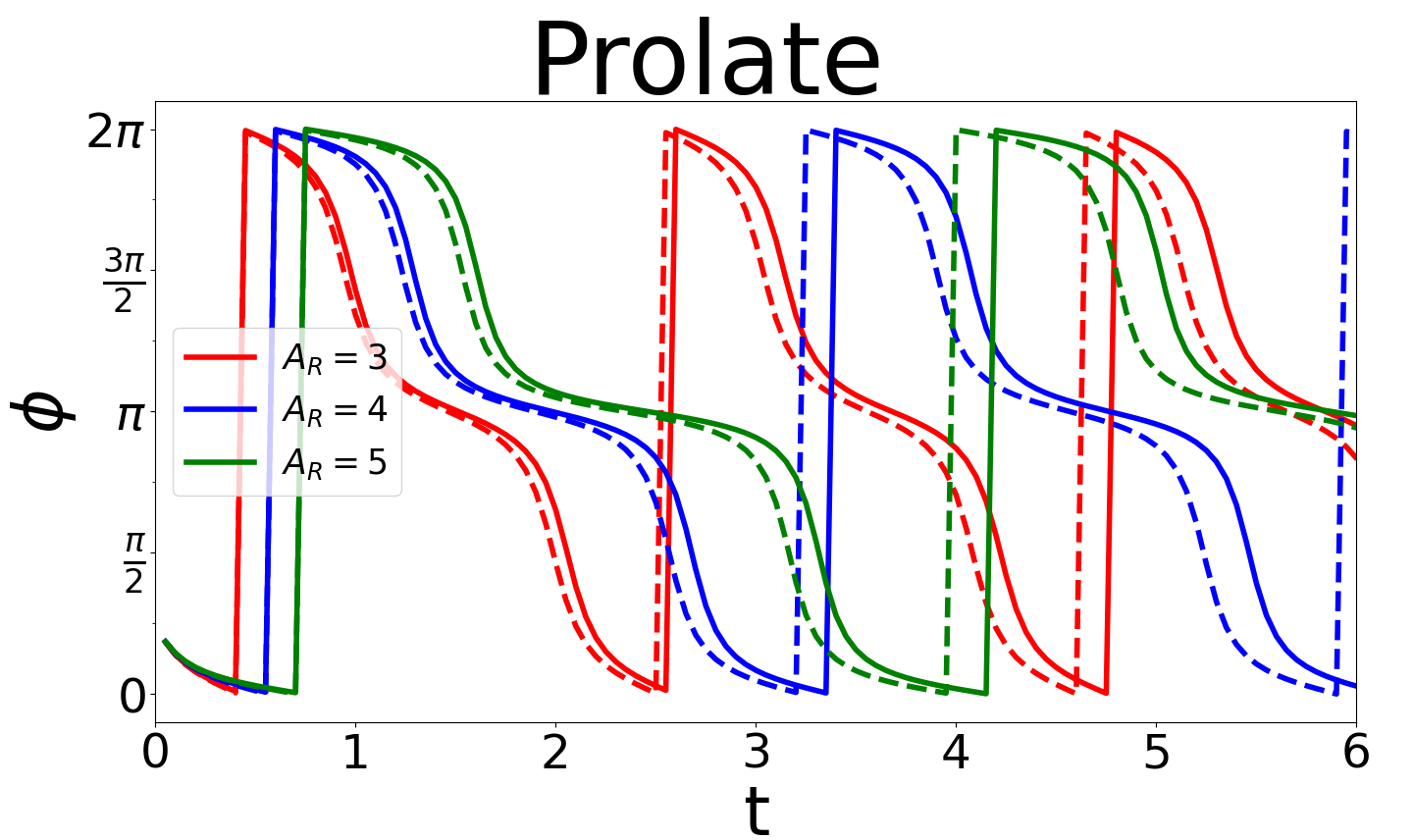}}
\subfloat[]{\includegraphics[width=0.45\linewidth]{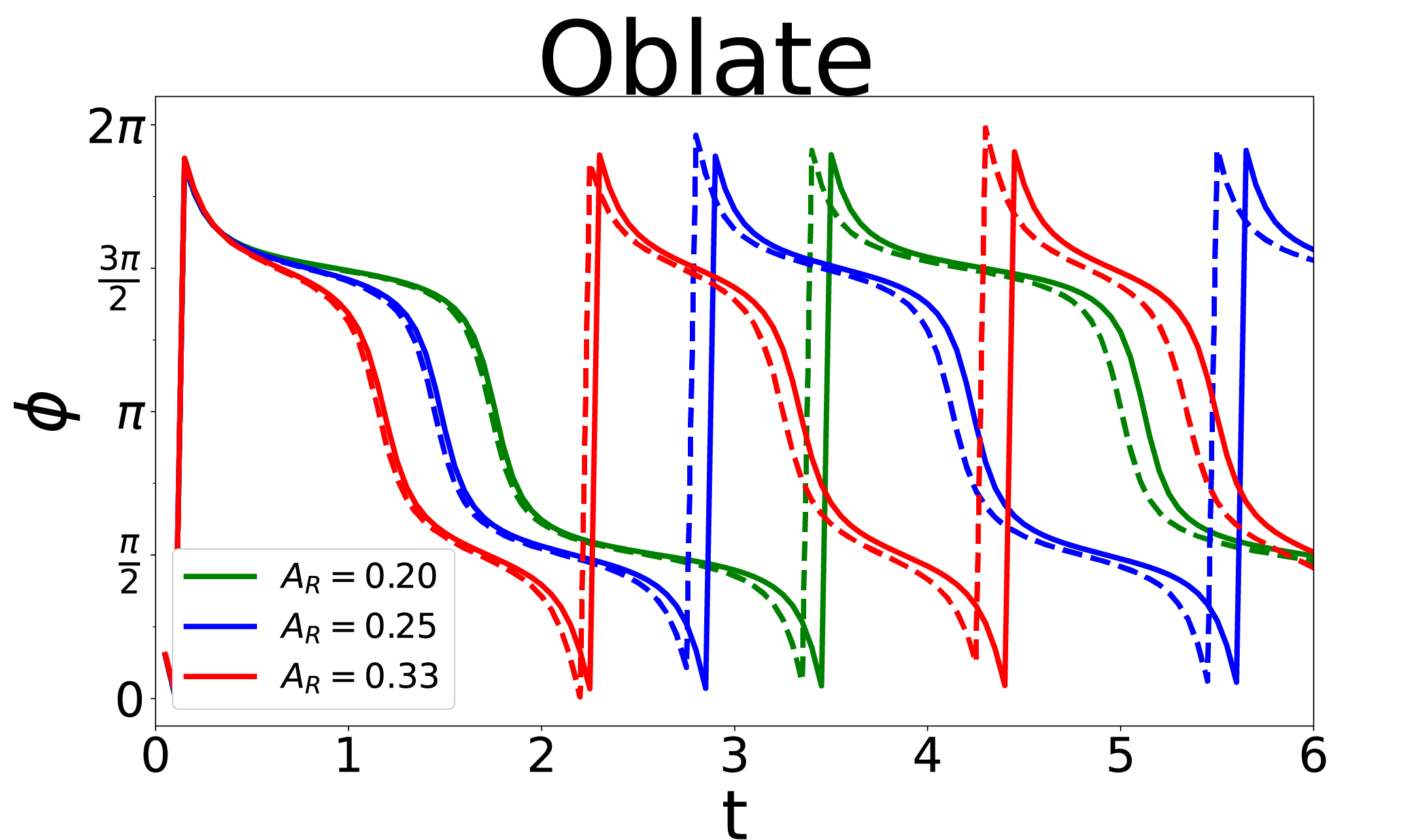}}
\hfill
\subfloat[]{\includegraphics[width=0.45\linewidth]{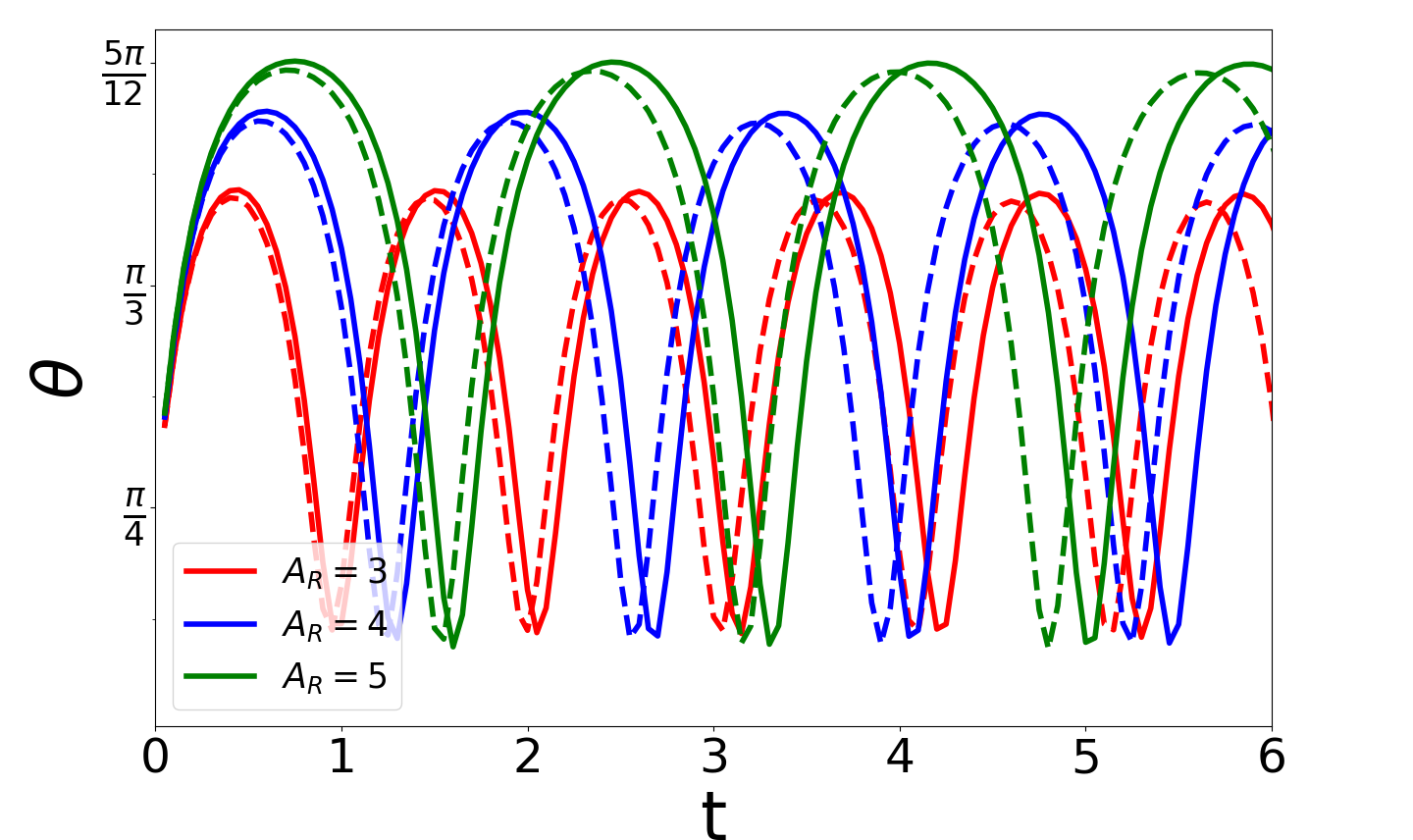}}
\subfloat[]{\includegraphics[width=0.45\linewidth]{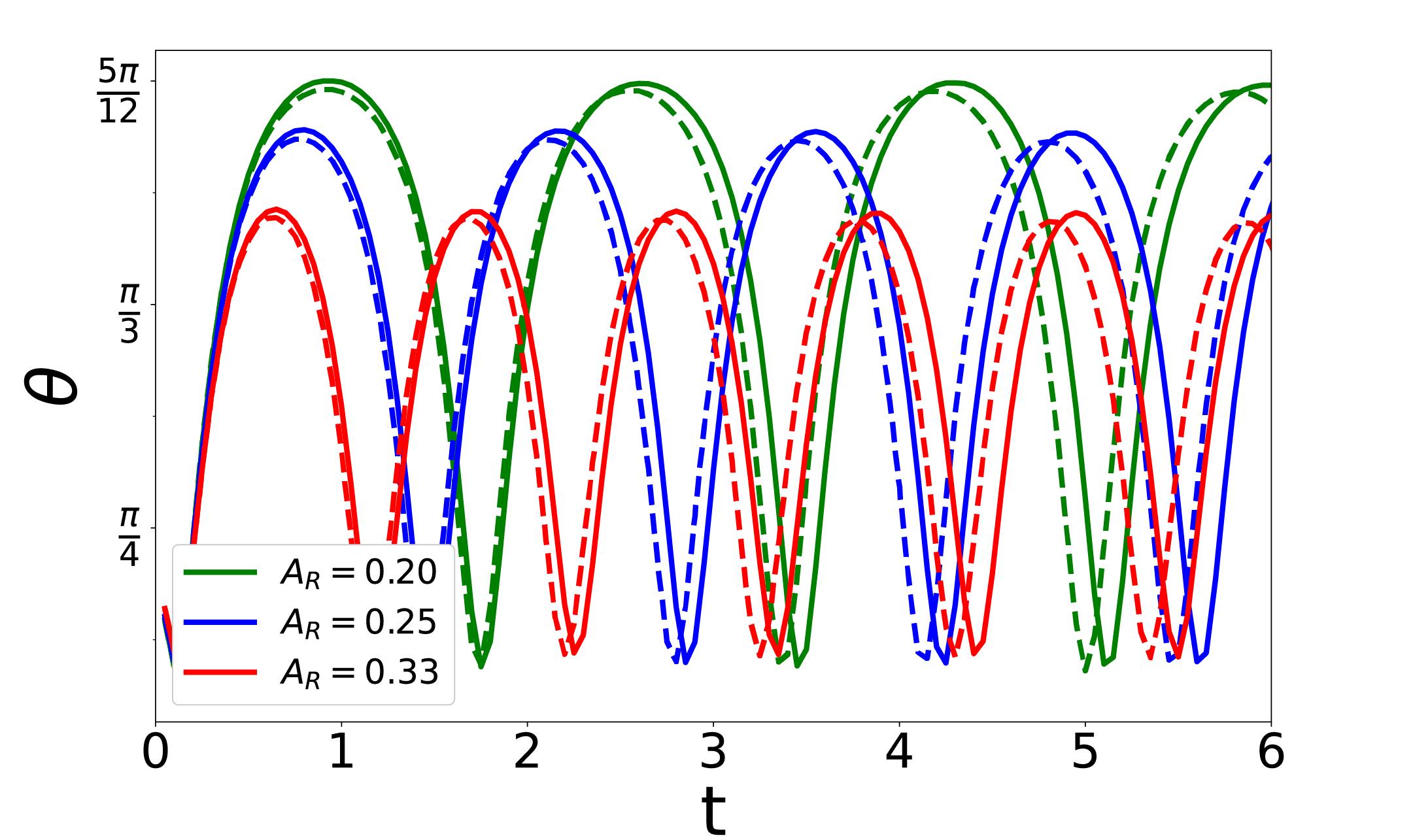}}
\caption{Effect of shape parameter ($A_R$) on the orientation trajectories of prolate and oblate spheroids in the large Carreau number limit $(Cu \gg 1)$.  The plots on the left show prolate spheroids and those on the right show oblate spheroids.  The solid lines denote shear thinning flows with $Cu =1000$ and $n =0.3$, while the dashed lines denote the Newtonian plateau ($Cu \rightarrow \infty$) with dimensionless viscosity $\beta = 0.1$.  For all the plots, the initial orientation is $\theta_0 =\pi/4,\phi_0 =\pi/4$. }
\label{fig:High_AR}
\end{figure}

Fig.~ \ref{fig:High_AR} plots the angle trajectories $(\phi(t), \theta(t))$ for oblate and prolate particles at the second Newtonian pleateau ($Cu \rightarrow \infty$ limit) and the shear-thinning region ($Cu \gg 1$ but finite).  Different values of shape parameter ($A_R$) are considered.  A few key observations can be made.  First, the figure shows that the time period and the amplitude for $\theta$ are smaller in the Newtonian plateau regime ($Cu \rightarrow \infty$) than the shear thinning regime ($Cu$ finite). The reason behind these trends is that the shear thinning regime has a larger viscosity than the Newtonian plateau, which leads to lower strain rates in the fluid, and hence larger time periods for tumbling and larger oscillation amplitude for $\theta$ as discussed previously.  The figure also shows that the further the particle shape deviates from a sphere ($A_R > 1$ for prolates and $A_R < 1$ for oblates), the difference in dynamics amplifies between the Newtonian plateau case and the shear thinning case.  Lastly, orientation-wise, the angle $\phi$ behaves similarly in both the large Carreau and the small Carreau number limits (compare Fig. \ref{fig:High_AR} with Fig. \ref{fig:Low_AR}).  In both cases, the spheroid likes to spend more time with the long axis aligned in the flow ($x$) direction, which corresponds to $\phi =m\pi$ for prolate particles and $\phi =(m+\frac{1}{2})\pi$ for oblate particles. In these orientations, the difference in shear thinning and Newtonian cases is minimum (see Fig.~\ref{fig:High_AR}(a,b)), while the orientations with the largest moment arm exhibit the largest differences between the shear thinning and Newtonian cases.  

\begin{figure}[t]
\subfloat[]
{\includegraphics[width=0.5\linewidth]{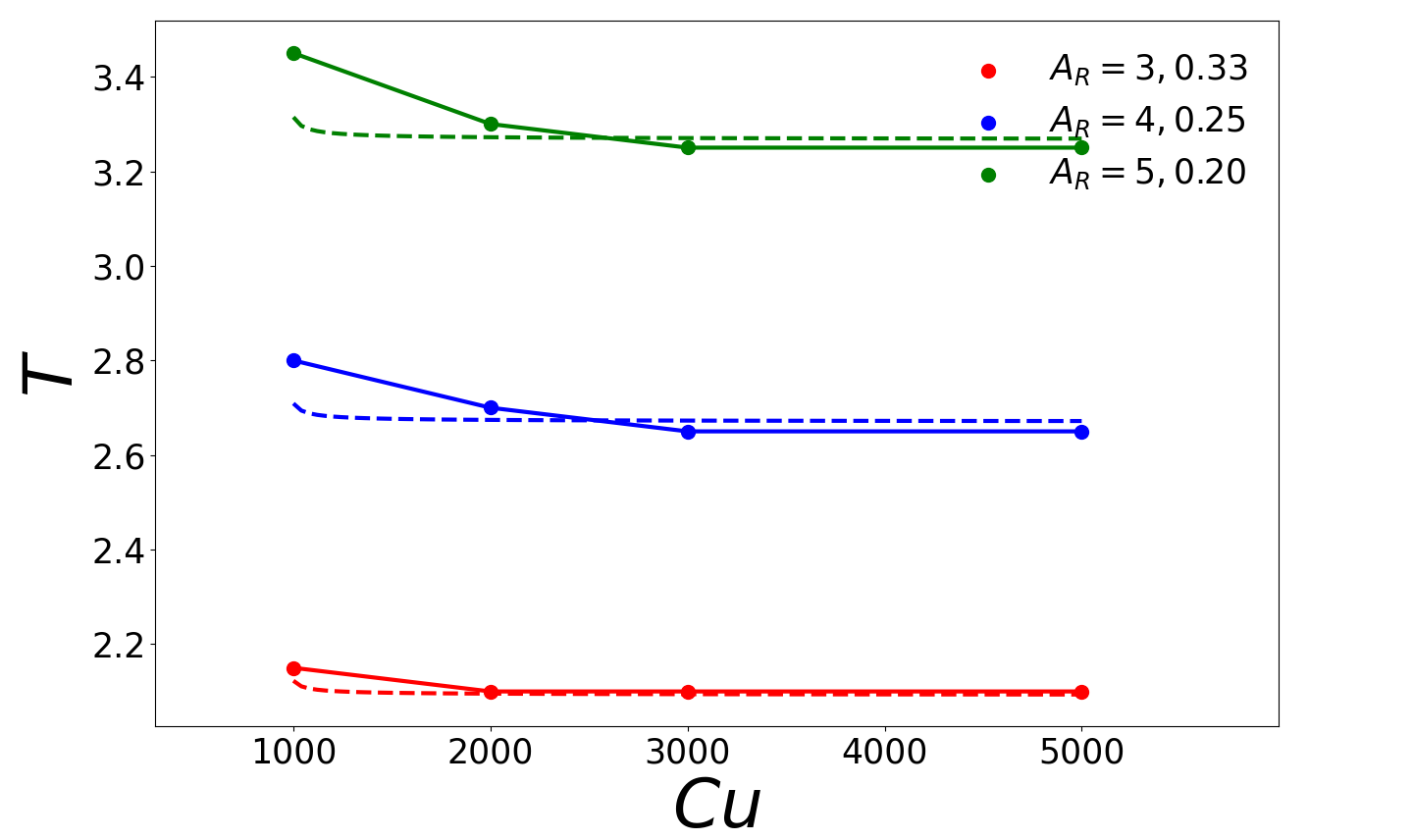}}
\subfloat[]
{\includegraphics[width=0.5\linewidth]{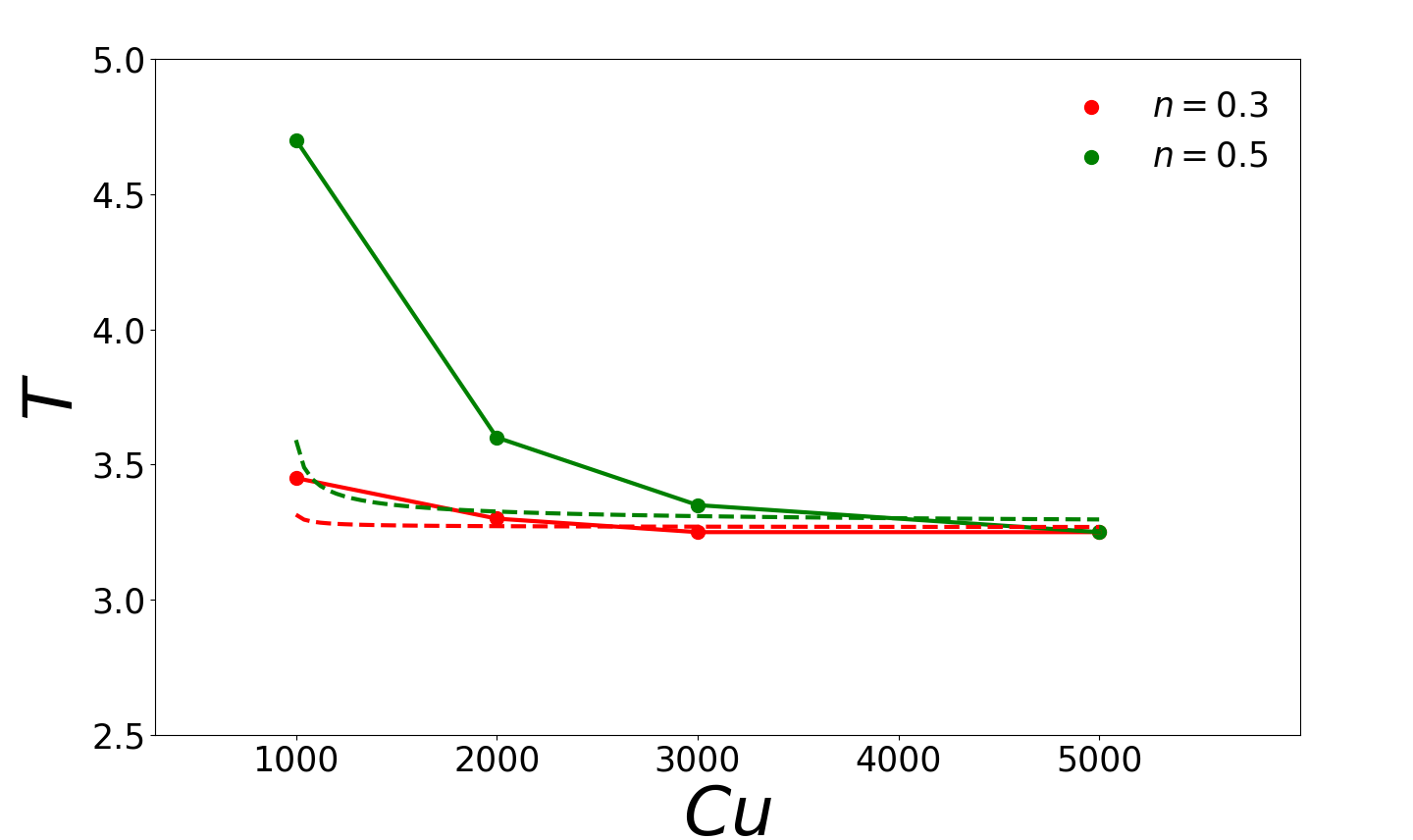}}

\caption{Tumbling time period for both prolate and oblate spheroids in the $Cu \gg 1$ limit for different values of (a) shape parameter $A_R$ and (b) power law index $n$. The initial orientation is ($\theta_0 =\pi/4,\phi_0 =\pi/4$) and the rheological parameter $\beta = 0.1$. For (a) $n =0.3$, and for (b) $A_R =5$. The solid lines show the results from the reciprocal theorem based numerical estimation while the dotted lines show the result using Jeffrey's formula using the local shear rate -- i.e.,  $T =2\pi(A_R +1/A_R)\dot{\gamma}_{\text{loc}}^{-1}$, where $\dot{\gamma}_{\text{loc}} =\frac{\partial}{\partial y} \left[u_x^{\infty,(0)}+Cu^{n-1} u_x^{\infty,(1)}\right]$.} 
\label{fig:High_AR_Time}
\end{figure}

Fig.~\ref{fig:High_AR_Time} plots the tumbling time period of spheroids in the large Carreau number regime for different values of the shape parameter ($A_R$), Carreau number ($Cu$), and power-law index ($n$).
These plots show that the time period decreases to a constant value as the Carreau number increases, reaching a Newtonian plateau independent of $Cu$ as $Cu \rightarrow \infty$.  The changes in the period however are much more modest than what was seen in the small Carreau number limit (compare Fig ~\ref{fig:Low_Time_AR_N} to Fig.~\ref{fig:High_AR_Time}).  The reason for this observation is that changes in the viscosity scale as $\epsilon (n-1) Cu^2$ in the small Carreau number limit, while the changes scale as $\epsilon Cu^{n-1}$ in the large Carreau number limit, which exhibits a weaker variation with respect to both $Cu$ and $n$.  Similar to what was observed before, we find the time period is the same for prolate and oblate particles of the same aspect ratio, and the effects of shear thinning are enhanced the further the particles deviate from a sphere. Lastly, in Fig.~\ref{fig:High_AR_Time} we also plot the period computed from the classic Jeffrey orbit equation $T = 2\pi (A_R + 1/A_R) \dot{\gamma}_{loc}^{-1}$, where $\dot{\gamma}_{loc}$ is the local shear rate at the center of mass of the particle.  One would expect that $\dot{\gamma}_{loc}$ will change with Carreau number ($Cu$) and power-law index ($n$), and thus explain some of the trends seen in the graph.  We see that while the equation does explain some of the trends, it does not overlap with the simulation results, although the agreement appears to be better than the small Carreau number limit in Fig \ref{fig:Low_Time_AR_N}.  Again, these results suggest that one cannot use simple ideas from Stokes flow to model the tumbling behavior of particles in shear thinning fluids.
The last point we would like to illustrate is that the time period in Fig.~\ref{fig:High_AR_Time} has different axes than plot in the small Carreau number limit (Fig. \ref{fig:Low_Time_AR_N}).  The tumbling period in the $Cu \rightarrow \infty$ limit is $\beta$ times smaller than the $Cu = 0$ limit, due to the fact that the shear rate in the channel is $1/\beta$ times larger.  This will play a role in the trends discussed next section when we compare the tumbling behavior of spheroids in pressure driven flows versus simple shear flow.

\section{Comparison with microhydrodynamics of spheroids in simple shear flow}
\label{sec:LinearFlow}
In this section, we explore the motion of a spheroid in a Couette flow of a shear thinning fluid. Such a flow is shown schematically in Fig.~\ref{fig:Flow_Linear_Schemtic}, where as before, the coordinate axes are placed in the center of the channel. However, here the flow is actuated by the top wall moving at a constant velocity $2\mathcal{U}_c$. We non-dimensionalize all lengths by height $h$, velocities by $\mathcal{U}_c$, shear rates by $\dot{\gamma}_c = \mathcal{U}_c/h$, times by $\dot{\gamma}_c^{-1}$, viscosities by $\eta_0$, and stresses by $\tau_c = \eta_0 \dot{\gamma}_c$.The  velocity field in the absence of the particle is given in dimensionless terms as:
\begin{figure}[t]
\centering
{\includegraphics[width=0.3\linewidth]{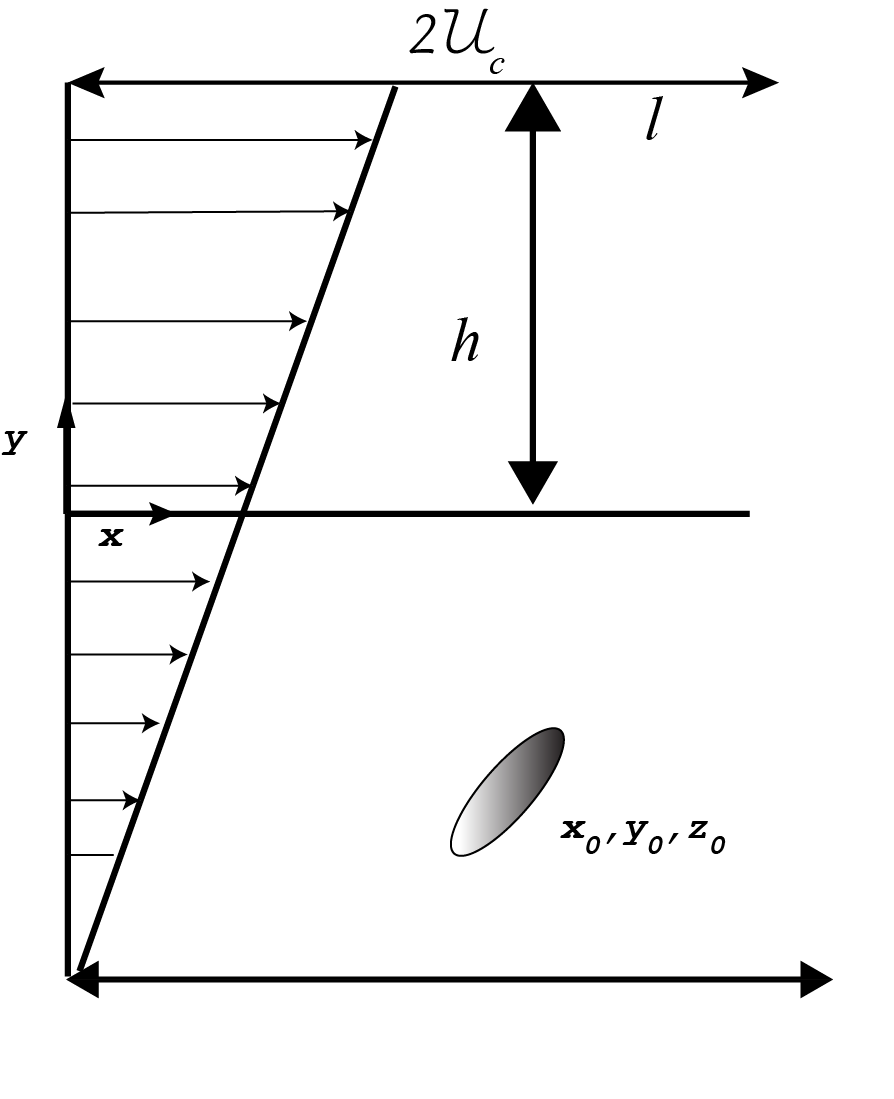}}
\caption{Spheroid in a Couette flow of a shear thinning fluid. The apparatus is same as that for pressure driven flow with the difference being that the flow is actuated by the motion of the top wall.}
\label{fig:Flow_Linear_Schemtic}
\end{figure}
\begin{equation}
     {u}_x^{\infty} = 1+y
\end{equation}
and the Carreau number is given by:
\begin{equation}
Cu = \lambda_t \dot{\gamma}_c = \lambda_t \mathcal{U}_c/h
\end{equation}
 We observe that the undisturbed velocity field is independent of viscosity, and is therefore same for both Newtonian as well as for the perturbative shear thinning cases ($\mathcal{O}(Cu^2)$  and $\mathcal{O}(Cu^{n-1}$)). 

The introduction of a spheroid into the linear background flow perpetuates a disturbance field around the particle, in addition to the undisturbed background flow. As before, to delineate the particle motion, we perturb the disturbance field in two limits: small Carreau number (with $Cu^2$ as the perturbation parameter) and large Carreau number (with $Cu^{n-1}$ as the perturbation parameter). In both these perturbative solution schemes, the leading order problem is Newtonian, whose solution has already been known in literature and is available in standard textbooks (\cite[Ch.~3]{Graham2018}) For the shear thinning correction, we appeal to the reciprocal theorem and solve for the particle's rigid body motion at
 $\mathcal{O}(Cu^2)$ for small Carreau number limit and at $\mathcal{O}(Cu^{n-1})$ for the large  Carreau number limit.

\begin{figure}[t]
\centering
\subfloat[]{\includegraphics[width=0.45\linewidth]{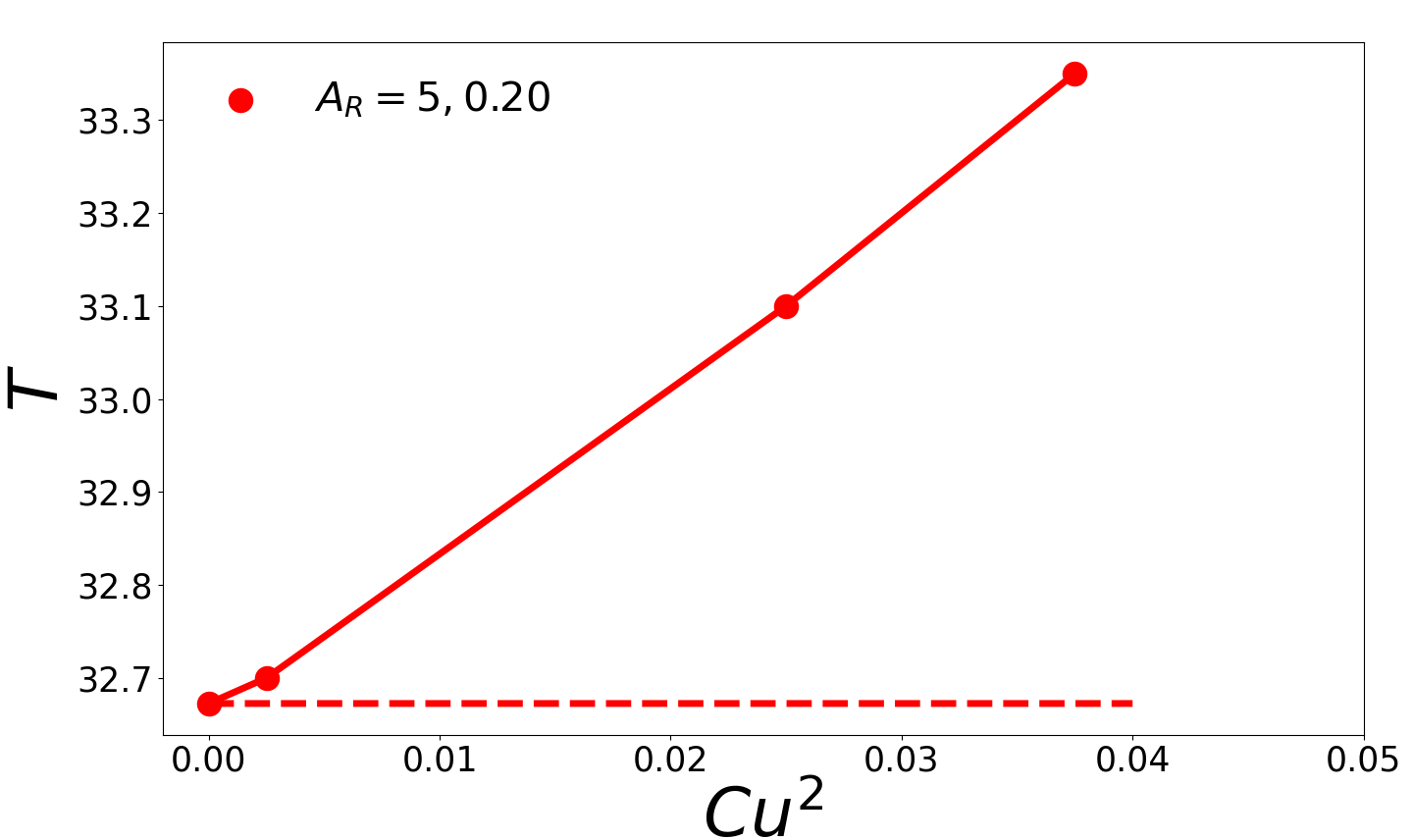}}
\subfloat[]{\includegraphics[width=0.45\linewidth]{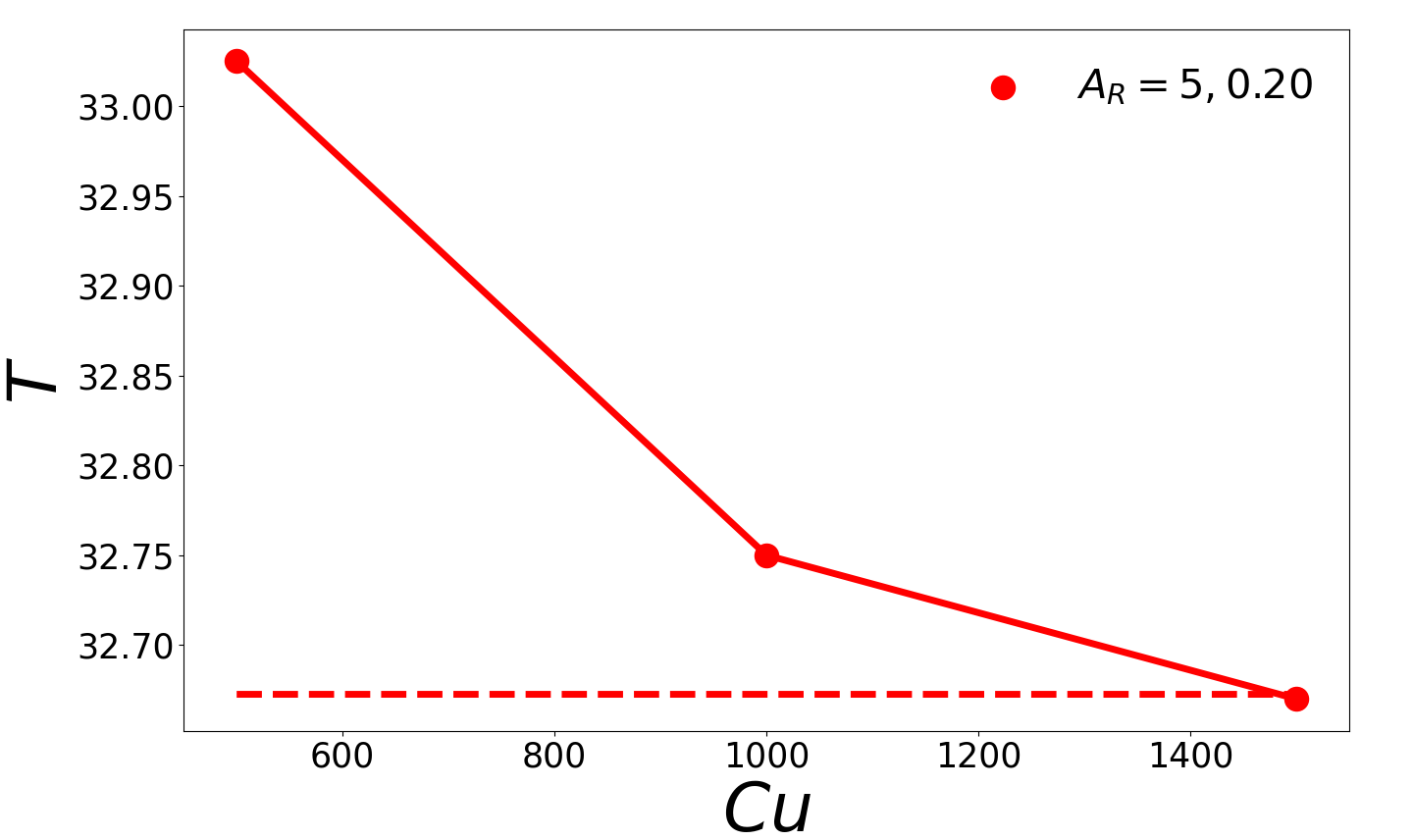}}

\caption{Time period of revolution of prolate and oblate spheroids in linear flows of shear thinning fluids in both (a) small $Cu$ and (b) large $Cu$ regimes. The solid curves show the shear thinning solution obtained using reciprocal theorem based simulations, while the dashed curves show the Newtonian solution using Jeffrey orbit formula. Rheological parameters include $n=0.3$ and $\beta =0.1$ while the initial orientation is given by $(\theta,\phi) =(\pi/4,\pi/4)$ }
\label{fig:Linear_Time}
\end{figure}

The result of this analysis leads us to the tumbling period of a spheroid in Couette flow of a  shear thinning fluid in both the small Carreau number and large Carreau number perturbative schemes. The results have been plotted in Fig. \ref{fig:Linear_Time}. Here, we observe that for the small $Cu$ case in Fig \ref{fig:Linear_Time}(a) , the time period increases with an increase in shear thinning. This result has also been reported earlier by \cite{Elfring_Main}. In the Newtonian limit of viscosity ($Cu \to 0$), the time period also reduces to that given by the Jefferey orbit formula (see dotted curve). More interesting for the current analysis is the result pertaining to the large $Cu$ number perturbative limit, as plotted in Fig.\ref{fig:Linear_Time}(b). \textit{This is a novel result not published before in literature.} Here, we observe that as the fluid around the particle continues to shear thin, the time period begins to decrease. At extremely large Carreau numbers, the time period then decreases back to its Jefferey orbit formula. We note that in the Newtonian limit of both these perturbative schemes ($Cu \to 0$ and $Cu \to \infty$), the time period reduces to the same constant value.

\begin{figure}[t]

{\includegraphics[width=0.5\linewidth]{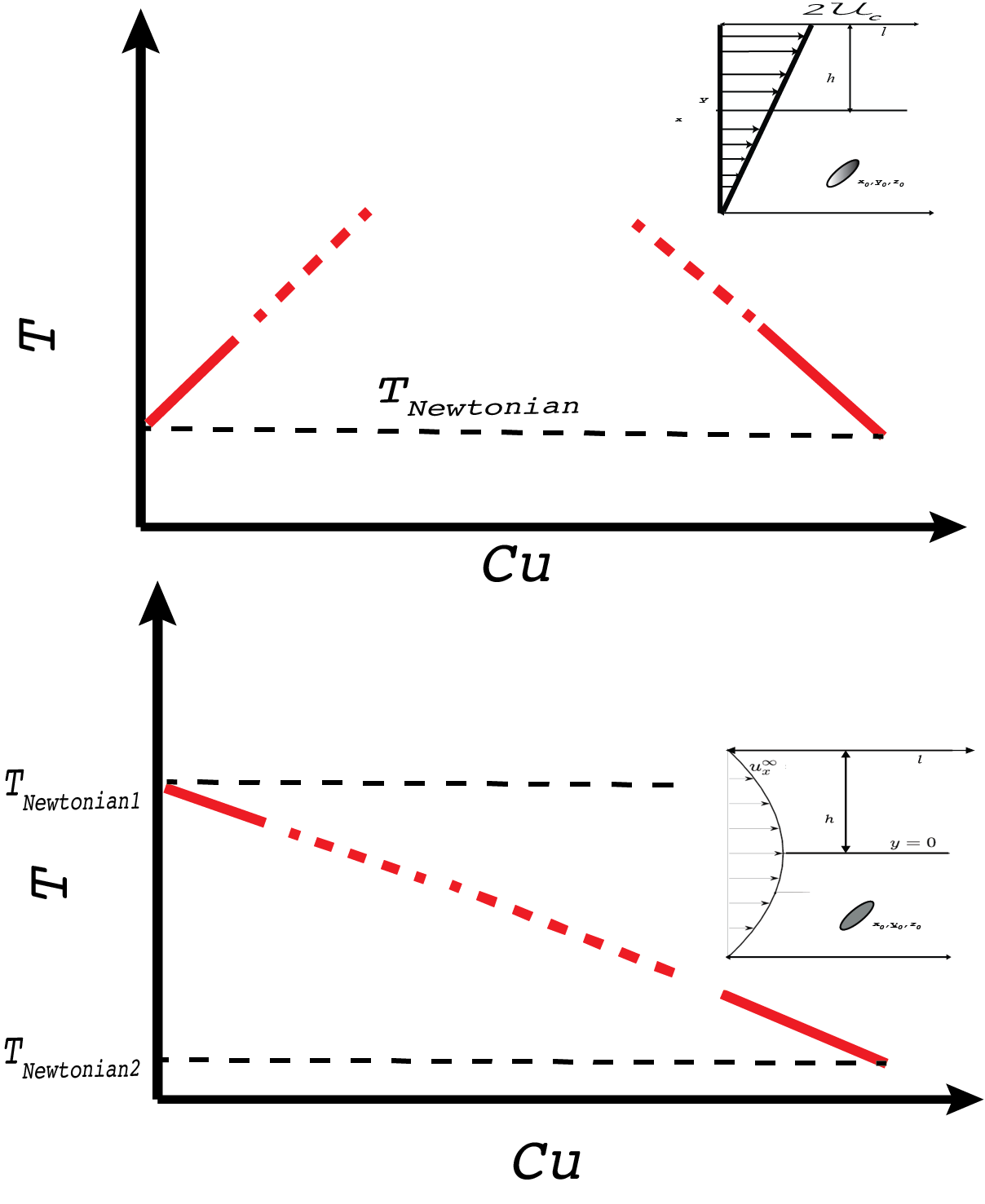}}

\caption{ Cartoon of the time period of revolution for spheroids in linear (top) and pressure driven (bottom) flows of shear thinning fluids (figure not to scale). The solid red curves illustrate known behavior from the perturbative analysis based on the reciprocal theorem, while the dotted red curves are hypothesis. The dotted black lines show the corresponding Newtonian plateaus. $T_{Newtonian2} = \beta \times T_{Newtonian1}$ }
\label{fig:Time_Linear_Schematic}
\end{figure}

In comparison to the pressure driven flow  as discussed in the previous sections, we notice the following differences in the microhydrodynamics of spheroids in linear flows of shear thinning fluids:
\begin{enumerate}
    \item In pressure driven flows, the time period decreases monotonically with an increase in Carreau number. On the other hand, in the case of linear flows, the time period shows a non-monotonic behavior. In the low Carreau number regime, the time period increases with increase in Carreau number; whilst on the other hand, in the high Carreau number case, the time period decreases with increases in Carreau number.
    \item In the pressure driven flows, the time period at the two Newtonian limits (low and high Carreau) are different  because the background flow and the Newtonian viscosity at these two limits are different. In the linear flow, the time period at the two Newtonian limits are the same, because the background flow at these two limits are also the same.   
\end{enumerate}

These two differences are schematically shown in Fig. \ref{fig:Time_Linear_Schematic}
\section{Conclusion}
\label{sec:conclusion}

In this paper, we analysed the motion of a spheroid in a pressure driven flow of a shear thinning fluid. The shear thinning rheology is captured by the Carreau model. We employ a perturbative approach in conjunction with the reciprocal theorem to delineate the orientational dynamics of prolate and oblate spheroids. First, an approximate version of the Carreau model is rendered, in both the small and large Carreau number limits. Next, the classical pressure driven (Poiseuille) flow equations are amended to take into account the shear thinning corrections due to rheology. Finally, using the reciprocal theorem and mobility relationships for ellipsoids in Newtonian flow, the orientational kinematics of the problem is evaluated numerically. 


We have the following conclusions to make:
\begin{itemize}
   \item Spheroids in shear thinning fluids tumble in a periodic fashion.  In other words, shear thinning does not resolve the degeneracy of Jeffrey’s orbits found in Newtonian fluids.  The degeneracy of Jefferey's orbits is attributed to the symmetry of the momentum equations for a Newtonian fluid. The shear thinning rheology, even though nonlinear, still preserves the symmetry of the momentum equations.  
   
   \item In the small Carreau number limit ($Cu \ll 1$), we find the spheroid's tumbling behavior to be very different in a pressure driven flow versus a linear shear flow.  In a pressure driven flow, shear thinning ($Cu > 0$) gives rise to a smaller tumbling time period compared to a Newtonian fluid ($Cu = 0$). When the particle is in a shear flow, shear thinning gives rise to the opposite trend -- i.e., a larger tumbling time period compared to its Newtonian counterpart \cite{Elfring_Main}.  The reason for these opposing trends is that in the pressure driven flow studied here, the flow is pressure controlled, while in a shear flow, the flow is kinematically controlled.  In the former case, shear thinning gives rise to a larger shear rate on the particle, and hence a faster tumbling period.  In the latter case, shear thinning gives rise to a lower stress on the particle, and hence a slower tumbling period. 
   
   \item We also observe a very different rotational behavior between spheroids in a shear thinning fluid and a viscoelastic fluid.  In a shear thinning fluid, we find that the rotational speed is the fastest for prolate particles  when they are aligned with the shear-gradient direction, and slowest when they are aligned with the flow direction.  The converse is true for prolate particles in a Boger fluid (e.g., second order fluid  \cite{Wang_Narsimhan_POF,Tai_Wang_Narsimhan_JFM}).  This tendency is attributed to the fact the hydrodynamic torque is due to shear stresses for the case of shear thinning flows while the torque is due to normal stresses for the case of viscoelastic flows. The different origins of the hydrodynamic torque thus implies that the torque's moment arm is the largest when the prolate particle is oriented along the flow direction for viscoelastic fluids, while the moment arm is the largest when the prolate particle is oriented along the shear direction for shear thinning fluids.  Similar trends are also seen for oblate particles as well.

    \item In a simple shear flow, the time period of tumbling of spheroids follow a non monotonic trend with respect to Carreau number. At small Carreau number, the time period increases from its Newtonian value, and then at high Carreau number it decreases back to the same Newtonian value. On the other hand, in the case of pressure driven flow, the time period of tumbling decreases monotonically with Carreau number. Moreover, the time period in the different Newtonian limits $(Cu \to 0)$ and $(Cu \to \infty)$ are different for pressure driven flows. The time period in the Newtonian limit of high Carreau $Cu \to \infty$ is $\beta$ times the time period in the Newtonian limit of low Carreau ($Cu \to 0$) for pressure driven flows.
    
\end{itemize}

In the future, the present analysis may be extended to incorporate the effects of other inelastic non Newtonian fluids like those with a spatially varying viscosity \cite{Datt_Elfring_Viscosity_Gradient,Dandekar_Ardekani_Swimming_Sheet_Viscosity} or with spatially varying density \cite{VMS22,VS2022,MA2022}. Similarly, instead of spheroids and ellipsoids, particles of other shapes like sheets \cite{Dandekar_Ardekani_Density_Sheet_JFM} , slender bodies \cite{SK06} and non fore-aft symmetric particles \cite{CM16} may be analysed with the same theoretical framework. Other types of flow problems beyond pressure driven flows, like sedimentation \cite{Kim86} for shear thinning fluids may be analysed. Finally, higher order effects like those of inertia \cite{MS18,SK06,DMS16}, particle size \cite{AS2023}, and effect of walls may also be taken into account.

\section*{Acknowledgments}
The authors acknowledge funding from the American Chemical Society Petroleum Research Fund (Grant No. ACS PRF 61266-DNI9).

\section*{Appendix A:  Derivation of reciprocal theorem for rigid body motion around particle} \label{sec:AppendixA}

Suppose we have two velocity fields in the same control volume $V$ of a liquid.  Both velocity fields satisfy the Stokes equations with a body force –  i.e.,
\begin{equation}
\frac{\partial u_i}{\partial x_i} = 0; \qquad \frac{\partial \tau_{ij}^N}{\partial x_j} + b_i = 0
\end{equation}
where $\tau_{ij}^N = \mu \dot{\gamma}_{ij} – p\delta_{ij}$ is the standard Newtonian stress tensor with viscosity $\mu$, and $b_i$ is a spatially varying body force.  We will add superscripts to the symbols above to demarcate the two flow fields -- $(\alpha)$ is flow field one, while $(\chi)$ is flow field two.  These flow fields are related to each other via Green’s second identity, which states that:
\begin{equation} \label{eq:Green_second_identity}
\int_S u_i^{(\alpha)}\tau_{ij}^{N, (\chi)} n_j dS + \int_V u_i^{(\alpha)} b_i^{(\chi)}dV = \int_S u_i^{(\chi)}\tau_{ij}^{N, (\alpha)} n_j dS + \int_V u_i^{(\chi)} b_i^{(\alpha)}dV
\end{equation}

In the above expression, $S$ is the surface of the control volume and $n_i$ is the outward pointing normal vector for the control volume.  While this expression appears esoteric, it is quite powerful.  It states that if one knows information about one flow field (e.g., flow ($\chi$)), one can obtain information about the other flow field (e.g., flow ($\alpha$)).  This expression can also be extended to non-Newtonian fluids, as will be illustrated below.

We will let the two flow fields for our problem be the following:

\paragraph{Flow 1:  Non-Newtonian flow around rigid particle}
We will choose flow one (with superscript $\alpha$) to be the disturbance flow around a particle with background velocity $u_i^{\infty}$ and body force $b_i = \frac{\partial }{\partial x_j} \left( \tau_{ij}^{ex}\right)$. Thus, we let:
\begin{subequations}
\begin{align}
u_i^{(\alpha)} &= u_i – u_i^{\infty}\\
\tau_{ij}^{N, (\alpha)} &= \tau_{ij}^N - \tau_{ij}^{N, \infty}\\
b_i^{(\alpha)} &= \frac{\partial}{\partial x_j} \left(\tau_{ij}^{ex} -  \tau_{ij}^{ex, \infty} \right)
\end{align}
\end{subequations}

In the above expression, quantities with the superscript ``$\infty$'' denote velocity and stress fields in the absence of the particle (i.e., due to $u_i^{\infty}$ only), while quantities without the superscript are fields with the particle present.  We choose disturbance quantities for convenience, as it will simplify the algebra later.

\paragraph{Flow 2:  Stokes flow around particle from rigid body motion}
We will let flow two (with superscript $\chi$) be Stokes flow around the same particle undergoing rigid body motion.  In other words, we let 
\begin{subequations}
\begin{align}
u_i^{(\chi)} &= v_i\\
\tau_{ij}^{N, (\chi)} &= \Sigma_{ij}\\
b_i^{(\chi)} &= 0
\end{align}
\end{subequations}
where $v_i$ and $\Sigma_{ij}$ are the velocity and stress fields around the particle from rigid body motion.  On the particle surface, $v_i = V_i +\epsilon_{ijk} \omega_j x_k$, where $x_k$ is the position vector from the particle’s center of mass, while $V_i$ and $\omega_i$ are the translational and rotational speeds.  The external force and torque on the particle will be $F_i^{aux}$ and $T_i^{aux}$.  These will be related to the translational and rotational speed through known resistance relationships.\\

Let us now substitute the information about the two flows into the integral expression \eqref{eq:Green_second_identity} above.  We will choose the control volume $V$ to be the volume outside of the particle.  Since we are dealing with disturbance quantities, we do not have to integrate over surfaces at infinity.  We obtain:
\begin{equation}
\int_{S_p} \left( u_i – u_i^{\infty} \right) \Sigma_{ij} n_j dS = \int_{S_p} v_i \left(\tau_{ij}^N - \tau_{ij}^{N,\infty} \right) n_j dS + \int_V v_i \frac{\partial}{\partial x_j} \left( \tau_{ij}^{ex} - \tau_{ij}^{ex, \infty} \right) dV
\end{equation}
where $S_p$ is particle surface and $n_i$ is the normal vector pointing into the particle (this is the outward pointing vector for the control volume $V$).  We can simplify the above expression using integration by parts on the last integral, noting that the total stress tensor is $\tau_{ij} = \tau_{ij}^N + \tau_{ij}^{ex}$.  This yields:

\begin{equation}
\int_{S_p} \left( u_i – u_i^{\infty} \right) \Sigma_{ij} n_j dS = \int_{S_p} v_i \left(\tau_{ij} - \tau_{ij}^{\infty} \right) n_j dS - \int_V \frac{\partial v_i}{\partial x_j} \left( \tau_{ij}^{ex} - \tau_{ij}^{ex, \infty} \right) dV
\end{equation}

The next step in the derivation is to note that the velocity fields are rigid body motion on the particle surface.  Thus, on the surface, $u_i = U_i + \epsilon_{ijk} \Omega_j x_k$, while for the other flow, $v_i = V_i + \epsilon_{ijk} \omega_j x_k$.  Substituting these expressions gives:

\begin{equation} \label{eq:reciprocal_integration}
U_i F_i^{aux} + \Omega_i T_i^{aux} = V_i F_i^{ext} + \omega_i T_i^{ext} + \int_{S_p} u_i^{\infty} \Sigma_{ij} n_j dS - \int_V \frac{\partial v_i}{\partial x_j} \left( \tau_{ij}^{ex} - \tau_{ij}^{ex, \infty} \right) dV
\end{equation}

In the above expression, $F_i^{ext}$ and $T_i^{ext}$ are the external force and torque on the particle in the non-Newtonian flow (i.e., flow $u_i$), while $F_i^{aux}$ and $T_i^{aux}$ are the external force and torque on the particle from the other flow (i.e., flow $v_i$).  

In the last part of the derivation, we note that all quantities associated with the flow field $v_i$ are linear in the rigid body translation and rotation ($V_i$, $\omega_i$).  In other words, we can write the flow field $v_i$ and stress field $\Sigma_{ij}$ in terms of these rigid body motions:

\begin{subequations} \label{eq:velocity_stress_aux}
\begin{align}
v_i &= v_{ik}^{trans} V_k + v_{ik}^{rot} \omega_k\\
\Sigma_{ij} &= \Sigma_{ijk}^{trans} V_k + \Sigma_{ijk}^{rot} \omega_k
\end{align}
\end{subequations}

In the above expression, $v_{ik}^{trans}$ is the flow field in the ``$i$'' direction due to unit translation in the ``$k$'' direction.  Similar notation follows for the other quantities.  For the force and torque $F_i^{aux}$ and $T_i^{aux}$, we write them as:

\begin{subequations} \label{eq:resistance_aux}
\begin{align}
F_i^{aux} = R^{FU}_{ij} V_j + R^{F\Omega}_{ij} \omega_j\\
T_i^{aux} = R^{TU}_{ij} V_j + R^{T\Omega}_{ij} \omega_j
\end{align}
\end{subequations}
where $R_{ij}^{FU}$, $R_{ij}^{F\Omega}$, $R_{ij}^{TU}$ and $R_{ij}^{T\Omega}$ are the Stokes-flow resistance tensors for the particle.  If we substitute the above two expressions \eqref{eq:velocity_stress_aux} and \eqref{eq:resistance_aux} into the integral expression \eqref{eq:reciprocal_integration} and perform some algebra, we obtain the final expression for the translational and rotational velocity of the particle in a non-Newtonian fluid.

\begin{equation}
    \begin{bmatrix}
R^{FU}_{ij} & R^{F\Omega}_{ij}\\
R^{TU}_{ij} & R^{T\Omega}_{ij}
\end{bmatrix}     \begin{bmatrix}
U_j \\
\Omega_j 
\end{bmatrix} = \begin{bmatrix}
F_i^{eff} \\
T_i^{eff} 
\end{bmatrix}
\end{equation}

The effective force and torque are given by equations \eqref{eq:effective_force_torque}-\eqref{eq:poly_force} in the main text.  

\section*{Appendix B:  Velocity fields from unit translation and rotation} \label{sec:AppendixB}
From Kim and Karilla \cite{KimKarilla2005}, the velocity fields $v_{ik}^{trans}$ and $v_{ik}^{rot}$ are given by the following expressions.  In these formulas, no summation is assumed for repeated indices unless explicitly stated.

\begin{subequations}
\begin{align}
    v_{ik}^{trans} = \frac{1}{16\pi\mu} R_{kk}^{FU} \left[ \delta_{ik} G_0 - x_k \frac{\partial G_0}{\partial x_i} + \frac{a_k^2}{2} \frac{\partial^2 G_1}{\partial x_i \partial x_k}\right] \\
    v_{ik}^{rot} = \frac{3}{32\pi\mu} R_{kk}^{T\Omega} \sum_{j=1}^3 \sum_{m = 1}^3 \epsilon_{jkm} \frac{\partial}{\partial x_m} \left[ \delta_{ij} G_1 - x_j \frac{\partial G_1}{\partial x_i} + \frac{a_j^2}{4} \frac{\partial^2 G_2}{\partial x_i \partial x_j}\right]
\end{align}
\end{subequations}

In the above expressions, the expression for $G_n$ is:

\begin{equation}
    G_n(x,y,z) = \int_{\lambda}^{\infty} \left( \frac{x^2}{a^2 + t} + \frac{y^2}{b^2 + t}  + \frac{z^2}{c^2 + t} - 1 \right)^n \frac{dt}{\Delta(t)}
\end{equation}
with $\Delta(t) = \sqrt{(a^2 + t)(b^2 + t)(c^2 + t)}$ and $\lambda(x,y,z)$ being the positive root of 

\begin{equation}
    \frac{x^2}{a^2 + t} + \frac{y^2}{b^2 + t}  + \frac{z^2}{c^2 + t} = 1
\end{equation}

\bibliographystyle{elsarticle-num-names}
\bibliography{references_microhydrodynamics.bib}

\end{document}